\documentclass[longauth]{aa}

\usepackage{graphicx}
\usepackage{natbib}
\usepackage{scalerel}
\usepackage{comment}

\usepackage{txfonts}
\usepackage{tabularx}

\usepackage[table]{xcolor}

\usepackage{txfonts}

\usepackage[pdfencoding=auto,psdextra]{hyperref}
\hypersetup{
    colorlinks=true,
    linkcolor=blue,
    filecolor=magenta,      
    urlcolor=blue,
    citecolor=blue
}
\makeatletter
\renewcommand*\aa@pageof{, page \thepage{} of \pageref*{LastPage}}
\makeatother

\usepackage[utf8]{inputenc}

\usepackage[switch, modulo]{lineno}

\renewcommand{\linenumbers}[0]{}

\usepackage{euclid}
\renewcommand{\farcs}{\mkern 2mu .\!\!^{\prime\prime}\mkern -1mu}
\renewcommand{\fdg}{\mkern 2mu .\!\!^\circ\mkern -1mu}

\begin{document}

    \title{\Euclid\/: Early Release Observations -- Internal kinematics and the convective-transition gap of NGC~6397\thanks{This paper is published on behalf of the Euclid Consortium}}
    \subtitle{High-precision multiple-pass photometry and astrometry}


\newcommand{\orcid}[1]{} 
\author{M.~Griggio\orcid{0000-0002-5060-1379}\thanks{\email{mgriggio@stsci.edu}}\inst{\ref{aff1}}
\and M.~Libralato\orcid{0000-0001-9673-7397}\inst{\ref{aff2}}
\and R.~Gerasimov\orcid{0000-0003-0398-639X}\inst{\ref{aff3}}
\and L.~R.~Bedin\orcid{0000-0003-4080-6466}\inst{\ref{aff2}}
\and A.~Bellini\orcid{0000-0003-3858-637X}\inst{\ref{aff1}}
\and J.~Anderson\orcid{0000-0003-2861-3995}\inst{\ref{aff1}}
\and E.~Dalessandro\orcid{0000-0003-4237-4601}\inst{\ref{aff4}}
\and E.~Vesperini\orcid{0000-0003-2742-6872}\inst{\ref{aff5}}
\and H.~Baumgardt\orcid{0000-0002-1959-6946}\inst{\ref{aff6}}
\and D.~Massari\orcid{0000-0001-8892-4301}\inst{\ref{aff4}}
\and M.~Cadelano\orcid{0000-0002-5038-3914}\inst{\ref{aff7},\ref{aff4}}
\and R.~E.~Ryan~Jr.~\orcid{0000-0003-0894-1588}\inst{\ref{aff1}}
\and I.~McDonald\orcid{0000-0003-0356-0655}\inst{\ref{aff8}}
\and F.~Annibali\inst{\ref{aff4}}
\and E.~Balbinot\orcid{0000-0002-1322-3153}\inst{\ref{aff9},\ref{aff10}}
\and J.-C.~Cuillandre\orcid{0000-0002-3263-8645}\inst{\ref{aff11}}
\and D.~Erkal\orcid{0000-0002-8448-5505}\inst{\ref{aff12}}
\and A.~M.~N.~Ferguson\inst{\ref{aff13}}
\and M.~Kluge\orcid{0000-0002-9618-2552}\inst{\ref{aff14}}
\and P.~B.~Kuzma\orcid{0000-0003-1980-8838}\inst{\ref{aff13}}
\and T.~Saifollahi\orcid{0000-0002-9554-7660}\inst{\ref{aff15},\ref{aff16}}
\and M.~Schirmer\orcid{0000-0003-2568-9994}\inst{\ref{aff17}}
\and K.~Voggel\orcid{0000-0001-6215-0950}\inst{\ref{aff15}}
\and B.~Altieri\orcid{0000-0003-3936-0284}\inst{\ref{aff18}}
\and S.~Andreon\orcid{0000-0002-2041-8784}\inst{\ref{aff19}}
\and N.~Auricchio\orcid{0000-0003-4444-8651}\inst{\ref{aff4}}
\and C.~Baccigalupi\orcid{0000-0002-8211-1630}\inst{\ref{aff20},\ref{aff21},\ref{aff22},\ref{aff23}}
\and M.~Baldi\orcid{0000-0003-4145-1943}\inst{\ref{aff24},\ref{aff4},\ref{aff25}}
\and A.~Balestra\orcid{0000-0002-6967-261X}\inst{\ref{aff2}}
\and S.~Bardelli\orcid{0000-0002-8900-0298}\inst{\ref{aff4}}
\and P.~Battaglia\orcid{0000-0002-7337-5909}\inst{\ref{aff4}}
\and A.~Biviano\orcid{0000-0002-0857-0732}\inst{\ref{aff21},\ref{aff20}}
\and E.~Branchini\orcid{0000-0002-0808-6908}\inst{\ref{aff26},\ref{aff27},\ref{aff19}}
\and M.~Brescia\orcid{0000-0001-9506-5680}\inst{\ref{aff28},\ref{aff29}}
\and S.~Camera\orcid{0000-0003-3399-3574}\inst{\ref{aff30},\ref{aff31},\ref{aff32}}
\and V.~Capobianco\orcid{0000-0002-3309-7692}\inst{\ref{aff32}}
\and C.~Carbone\orcid{0000-0003-0125-3563}\inst{\ref{aff33}}
\and J.~Carretero\orcid{0000-0002-3130-0204}\inst{\ref{aff34},\ref{aff35}}
\and M.~Castellano\orcid{0000-0001-9875-8263}\inst{\ref{aff36}}
\and G.~Castignani\orcid{0000-0001-6831-0687}\inst{\ref{aff4}}
\and S.~Cavuoti\orcid{0000-0002-3787-4196}\inst{\ref{aff29},\ref{aff37}}
\and A.~Cimatti\inst{\ref{aff7}}
\and C.~Colodro-Conde\inst{\ref{aff38}}
\and G.~Congedo\orcid{0000-0003-2508-0046}\inst{\ref{aff13}}
\and C.~J.~Conselice\orcid{0000-0003-1949-7638}\inst{\ref{aff8}}
\and L.~Conversi\orcid{0000-0002-6710-8476}\inst{\ref{aff39},\ref{aff18}}
\and Y.~Copin\orcid{0000-0002-5317-7518}\inst{\ref{aff40}}
\and F.~Courbin\orcid{0000-0003-0758-6510}\inst{\ref{aff41},\ref{aff42},\ref{aff43}}
\and H.~M.~Courtois\orcid{0000-0003-0509-1776}\inst{\ref{aff44}}
\and M.~Cropper\orcid{0000-0003-4571-9468}\inst{\ref{aff45}}
\and G.~De~Lucia\orcid{0000-0002-6220-9104}\inst{\ref{aff21}}
\and F.~Dubath\orcid{0000-0002-6533-2810}\inst{\ref{aff46}}
\and X.~Dupac\inst{\ref{aff18}}
\and M.~Farina\orcid{0000-0002-3089-7846}\inst{\ref{aff47}}
\and R.~Farinelli\inst{\ref{aff4}}
\and S.~Ferriol\inst{\ref{aff40}}
\and E.~Franceschi\orcid{0000-0002-0585-6591}\inst{\ref{aff4}}
\and S.~Galeotta\orcid{0000-0002-3748-5115}\inst{\ref{aff21}}
\and K.~George\orcid{0000-0002-1734-8455}\inst{\ref{aff48}}
\and C.~Giocoli\orcid{0000-0002-9590-7961}\inst{\ref{aff4},\ref{aff25}}
\and J.~Gracia-Carpio\orcid{0000-0003-4689-3134}\inst{\ref{aff14}}
\and A.~Grazian\orcid{0000-0002-5688-0663}\inst{\ref{aff2}}
\and F.~Grupp\inst{\ref{aff14},\ref{aff49}}
\and S.~V.~H.~Haugan\orcid{0000-0001-9648-7260}\inst{\ref{aff50}}
\and H.~Hoekstra\orcid{0000-0002-0641-3231}\inst{\ref{aff10}}
\and W.~Holmes\orcid{0009-0007-8554-4646}\inst{\ref{aff51}}
\and F.~Hormuth\inst{\ref{aff52}}
\and A.~Hornstrup\orcid{0000-0002-3363-0936}\inst{\ref{aff53},\ref{aff54}}
\and K.~Jahnke\orcid{0000-0003-3804-2137}\inst{\ref{aff17}}
\and M.~Jhabvala\inst{\ref{aff55}}
\and S.~Kermiche\orcid{0000-0002-0302-5735}\inst{\ref{aff56}}
\and A.~Kiessling\orcid{0000-0002-2590-1273}\inst{\ref{aff51}}
\and B.~Kubik\orcid{0009-0006-5823-4880}\inst{\ref{aff40}}
\and M.~K\"ummel\orcid{0000-0003-2791-2117}\inst{\ref{aff49}}
\and H.~Kurki-Suonio\orcid{0000-0002-4618-3063}\inst{\ref{aff57},\ref{aff58}}
\and A.~M.~C.~Le~Brun\orcid{0000-0002-0936-4594}\inst{\ref{aff59}}
\and S.~Ligori\orcid{0000-0003-4172-4606}\inst{\ref{aff32}}
\and P.~B.~Lilje\orcid{0000-0003-4324-7794}\inst{\ref{aff50}}
\and V.~Lindholm\orcid{0000-0003-2317-5471}\inst{\ref{aff57},\ref{aff58}}
\and I.~Lloro\orcid{0000-0001-5966-1434}\inst{\ref{aff60}}
\and G.~Mainetti\orcid{0000-0003-2384-2377}\inst{\ref{aff61}}
\and O.~Marggraf\orcid{0000-0001-7242-3852}\inst{\ref{aff62}}
\and M.~Martinelli\orcid{0000-0002-6943-7732}\inst{\ref{aff36},\ref{aff63}}
\and N.~Martinet\orcid{0000-0003-2786-7790}\inst{\ref{aff64}}
\and F.~Marulli\orcid{0000-0002-8850-0303}\inst{\ref{aff65},\ref{aff4},\ref{aff25}}
\and R.~J.~Massey\orcid{0000-0002-6085-3780}\inst{\ref{aff66}}
\and E.~Medinaceli\orcid{0000-0002-4040-7783}\inst{\ref{aff4}}
\and S.~Mei\orcid{0000-0002-2849-559X}\inst{\ref{aff67},\ref{aff68}}
\and M.~Meneghetti\orcid{0000-0003-1225-7084}\inst{\ref{aff4},\ref{aff25}}
\and E.~Merlin\orcid{0000-0001-6870-8900}\inst{\ref{aff2}}
\and G.~Meylan\orcid{0000-0001-6503-0209}\inst{\ref{aff69}}
\and A.~Mora\orcid{0000-0002-1922-8529}\inst{\ref{aff70}}
\and M.~Moresco\orcid{0000-0002-7616-7136}\inst{\ref{aff65},\ref{aff4}}
\and L.~Moscardini\orcid{0000-0002-3473-6716}\inst{\ref{aff65},\ref{aff4},\ref{aff25}}
\and R.~Nakajima\orcid{0009-0009-1213-7040}\inst{\ref{aff62}}
\and S.-M.~Niemi\orcid{0009-0005-0247-0086}\inst{\ref{aff71}}
\and C.~Padilla\orcid{0000-0001-7951-0166}\inst{\ref{aff72}}
\and S.~Paltani\orcid{0000-0002-8108-9179}\inst{\ref{aff46}}
\and F.~Pasian\orcid{0000-0002-4869-3227}\inst{\ref{aff21}}
\and K.~Pedersen\inst{\ref{aff73}}
\and W.~J.~Percival\orcid{0000-0002-0644-5727}\inst{\ref{aff74},\ref{aff75},\ref{aff76}}
\and V.~Pettorino\orcid{0000-0002-4203-9320}\inst{\ref{aff71}}
\and G.~Polenta\orcid{0000-0003-4067-9196}\inst{\ref{aff77}}
\and M.~Poncet\inst{\ref{aff16}}
\and L.~A.~Popa\inst{\ref{aff78}}
\and F.~Raison\orcid{0000-0002-7819-6918}\inst{\ref{aff14}}
\and A.~Renzi\orcid{0000-0001-9856-1970}\inst{\ref{aff79},\ref{aff80},\ref{aff4}}
\and J.~Rhodes\orcid{0000-0002-4485-8549}\inst{\ref{aff51}}
\and G.~Riccio\inst{\ref{aff29}}
\and E.~Romelli\orcid{0000-0003-3069-9222}\inst{\ref{aff21}}
\and M.~Roncarelli\orcid{0000-0001-9587-7822}\inst{\ref{aff4}}
\and R.~Saglia\orcid{0000-0003-0378-7032}\inst{\ref{aff49},\ref{aff14}}
\and Z.~Sakr\orcid{0000-0002-4823-3757}\inst{\ref{aff81},\ref{aff82},\ref{aff83}}
\and D.~Sapone\orcid{0000-0001-7089-4503}\inst{\ref{aff84}}
\and B.~Sartoris\orcid{0000-0003-1337-5269}\inst{\ref{aff49},\ref{aff21}}
\and P.~Schneider\orcid{0000-0001-8561-2679}\inst{\ref{aff62}}
\and A.~Secroun\orcid{0000-0003-0505-3710}\inst{\ref{aff56}}
\and P.~Simon\inst{\ref{aff62}}
\and C.~Sirignano\orcid{0000-0002-0995-7146}\inst{\ref{aff79},\ref{aff80}}
\and G.~Sirri\orcid{0000-0003-2626-2853}\inst{\ref{aff25}}
\and L.~Stanco\orcid{0000-0002-9706-5104}\inst{\ref{aff80}}
\and P.~Tallada-Cresp\'{i}\orcid{0000-0002-1336-8328}\inst{\ref{aff34},\ref{aff35}}
\and S.~Toft\orcid{0000-0003-3631-7176}\inst{\ref{aff85},\ref{aff86}}
\and R.~Toledo-Moreo\orcid{0000-0002-2997-4859}\inst{\ref{aff87},\ref{aff88}}
\and F.~Torradeflot\orcid{0000-0003-1160-1517}\inst{\ref{aff35},\ref{aff34}}
\and I.~Tutusaus\orcid{0000-0002-3199-0399}\inst{\ref{aff89},\ref{aff90},\ref{aff82}}
\and J.~Valiviita\orcid{0000-0001-6225-3693}\inst{\ref{aff57},\ref{aff58}}
\and T.~Vassallo\orcid{0000-0001-6512-6358}\inst{\ref{aff21},\ref{aff48}}
\and Y.~Wang\orcid{0000-0002-4749-2984}\inst{\ref{aff91}}
\and J.~Weller\orcid{0000-0002-8282-2010}\inst{\ref{aff49},\ref{aff14}}
\and F.~M.~Zerbi\orcid{0000-0002-9996-973X}\inst{\ref{aff19}}
\and E.~Zucca\orcid{0000-0002-5845-8132}\inst{\ref{aff4}}
\and V.~Scottez\orcid{0009-0008-3864-940X}\inst{\ref{aff92},\ref{aff93}}}
										   
\institute{Space Telescope Science Institute, 3700 San Martin Dr, Baltimore, MD 21218, USA\label{aff1}
\and
INAF-Osservatorio Astronomico di Padova, Via dell'Osservatorio 5, 35122 Padova, Italy\label{aff2}
\and
Department of Physics and Astronomy, University of Notre Dame, Nieuwland Science Hall, Notre Dame, 46556, Indiana, USA\label{aff3}
\and
INAF-Osservatorio di Astrofisica e Scienza dello Spazio di Bologna, Via Piero Gobetti 93/3, 40129 Bologna, Italy\label{aff4}
\and
Department of Astronomy, Indiana University, 727 East Third Street, Bloomington, IN 47405, USA\label{aff5}
\and
School of Mathematics and Physics, The University of Queensland, St. Lucia, QLD 4072, Australia\label{aff6}
\and
Dipartimento di Fisica e Astronomia "Augusto Righi" - Alma Mater Studiorum Universit\`a di Bologna, Viale Berti Pichat 6/2, 40127 Bologna, Italy\label{aff7}
\and
Jodrell Bank Centre for Astrophysics, Department of Physics and Astronomy, University of Manchester, Oxford Road, Manchester M13 9PL, UK\label{aff8}
\and
Kapteyn Astronomical Institute, University of Groningen, PO Box 800, 9700 AV Groningen, The Netherlands\label{aff9}
\and
Leiden Observatory, Leiden University, Einsteinweg 55, 2333 CC Leiden, The Netherlands\label{aff10}
\and
Universit\'e Paris-Saclay, Universit\'e Paris Cit\'e, CEA, CNRS, AIM, 91191, Gif-sur-Yvette, France\label{aff11}
\and
School of Mathematics and Physics, University of Surrey, Guildford, Surrey, GU2 7XH, UK\label{aff12}
\and
Institute for Astronomy, University of Edinburgh, Royal Observatory, Blackford Hill, Edinburgh EH9 3HJ, UK\label{aff13}
\and
Max Planck Institute for Extraterrestrial Physics, Giessenbachstr. 1, 85748 Garching, Germany\label{aff14}
\and
Universit\'e de Strasbourg, CNRS, Observatoire astronomique de Strasbourg, UMR 7550, 67000 Strasbourg, France\label{aff15}
\and
Centre National d'Etudes Spatiales -- Centre spatial de Toulouse, 18 avenue Edouard Belin, 31401 Toulouse Cedex 9, France\label{aff16}
\and
Max-Planck-Institut f\"ur Astronomie, K\"onigstuhl 17, 69117 Heidelberg, Germany\label{aff17}
\and
ESAC/ESA, Camino Bajo del Castillo, s/n., Urb. Villafranca del Castillo, 28692 Villanueva de la Ca\~nada, Madrid, Spain\label{aff18}
\and
INAF-Osservatorio Astronomico di Brera, Via Brera 28, 20122 Milano, Italy\label{aff19}
\and
IFPU, Institute for Fundamental Physics of the Universe, via Beirut 2, 34151 Trieste, Italy\label{aff20}
\and
INAF-Osservatorio Astronomico di Trieste, Via G. B. Tiepolo 11, 34143 Trieste, Italy\label{aff21}
\and
INFN, Sezione di Trieste, Via Valerio 2, 34127 Trieste TS, Italy\label{aff22}
\and
SISSA, International School for Advanced Studies, Via Bonomea 265, 34136 Trieste TS, Italy\label{aff23}
\and
Dipartimento di Fisica e Astronomia, Universit\`a di Bologna, Via Gobetti 93/2, 40129 Bologna, Italy\label{aff24}
\and
INFN-Sezione di Bologna, Viale Berti Pichat 6/2, 40127 Bologna, Italy\label{aff25}
\and
Dipartimento di Fisica, Universit\`a di Genova, Via Dodecaneso 33, 16146, Genova, Italy\label{aff26}
\and
INFN-Sezione di Genova, Via Dodecaneso 33, 16146, Genova, Italy\label{aff27}
\and
Department of Physics "E. Pancini", University Federico II, Via Cinthia 6, 80126, Napoli, Italy\label{aff28}
\and
INAF-Osservatorio Astronomico di Capodimonte, Via Moiariello 16, 80131 Napoli, Italy\label{aff29}
\and
Dipartimento di Fisica, Universit\`a degli Studi di Torino, Via P. Giuria 1, 10125 Torino, Italy\label{aff30}
\and
INFN-Sezione di Torino, Via P. Giuria 1, 10125 Torino, Italy\label{aff31}
\and
INAF-Osservatorio Astrofisico di Torino, Via Osservatorio 20, 10025 Pino Torinese (TO), Italy\label{aff32}
\and
INAF-IASF Milano, Via Alfonso Corti 12, 20133 Milano, Italy\label{aff33}
\and
Centro de Investigaciones Energ\'eticas, Medioambientales y Tecnol\'ogicas (CIEMAT), Avenida Complutense 40, 28040 Madrid, Spain\label{aff34}
\and
Port d'Informaci\'{o} Cient\'{i}fica, Campus UAB, C. Albareda s/n, 08193 Bellaterra (Barcelona), Spain\label{aff35}
\and
INAF-Osservatorio Astronomico di Roma, Via Frascati 33, 00078 Monteporzio Catone, Italy\label{aff36}
\and
INFN section of Naples, Via Cinthia 6, 80126, Napoli, Italy\label{aff37}
\and
Instituto de Astrof\'{\i}sica de Canarias, E-38205 La Laguna, Tenerife, Spain\label{aff38}
\and
European Space Agency/ESRIN, Largo Galileo Galilei 1, 00044 Frascati, Roma, Italy\label{aff39}
\and
Universit\'e Claude Bernard Lyon 1, CNRS/IN2P3, IP2I Lyon, UMR 5822, Villeurbanne, F-69100, France\label{aff40}
\and
Institut de Ci\`{e}ncies del Cosmos (ICCUB), Universitat de Barcelona (IEEC-UB), Mart\'{i} i Franqu\`{e}s 1, 08028 Barcelona, Spain\label{aff41}
\and
Instituci\'o Catalana de Recerca i Estudis Avan\c{c}ats (ICREA), Passeig de Llu\'{\i}s Companys 23, 08010 Barcelona, Spain\label{aff42}
\and
Institut de Ciencies de l'Espai (IEEC-CSIC), Campus UAB, Carrer de Can Magrans, s/n Cerdanyola del Vall\'es, 08193 Barcelona, Spain\label{aff43}
\and
UCB Lyon 1, CNRS/IN2P3, IUF, IP2I Lyon, 4 rue Enrico Fermi, 69622 Villeurbanne, France\label{aff44}
\and
Mullard Space Science Laboratory, University College London, Holmbury St Mary, Dorking, Surrey RH5 6NT, UK\label{aff45}
\and
Department of Astronomy, University of Geneva, ch. d'Ecogia 16, 1290 Versoix, Switzerland\label{aff46}
\and
INAF-Istituto di Astrofisica e Planetologia Spaziali, via del Fosso del Cavaliere, 100, 00100 Roma, Italy\label{aff47}
\and
University Observatory, LMU Faculty of Physics, Scheinerstr.~1, 81679 Munich, Germany\label{aff48}
\and
Universit\"ats-Sternwarte M\"unchen, Fakult\"at f\"ur Physik, Ludwig-Maximilians-Universit\"at M\"unchen, Scheinerstr.~1, 81679 M\"unchen, Germany\label{aff49}
\and
Institute of Theoretical Astrophysics, University of Oslo, P.O. Box 1029 Blindern, 0315 Oslo, Norway\label{aff50}
\and
Jet Propulsion Laboratory, California Institute of Technology, 4800 Oak Grove Drive, Pasadena, CA, 91109, USA\label{aff51}
\and
Felix Hormuth Engineering, Goethestr. 17, 69181 Leimen, Germany\label{aff52}
\and
Technical University of Denmark, Elektrovej 327, 2800 Kgs. Lyngby, Denmark\label{aff53}
\and
Cosmic Dawn Center (DAWN), Denmark\label{aff54}
\and
NASA Goddard Space Flight Center, Greenbelt, MD 20771, USA\label{aff55}
\and
Aix-Marseille Universit\'e, CNRS/IN2P3, CPPM, Marseille, France\label{aff56}
\and
Department of Physics, P.O. Box 64, University of Helsinki, 00014 Helsinki, Finland\label{aff57}
\and
Helsinki Institute of Physics, Gustaf H{\"a}llstr{\"o}min katu 2, University of Helsinki, 00014 Helsinki, Finland\label{aff58}
\and
Laboratoire d'etude de l'Univers et des phenomenes eXtremes, Observatoire de Paris, Universit\'e PSL, Sorbonne Universit\'e, CNRS, 92190 Meudon, France\label{aff59}
\and
SKAO, Jodrell Bank, Lower Withington, Macclesfield SK11 9FT, UK\label{aff60}
\and
Centre de Calcul de l'IN2P3/CNRS, 21 avenue Pierre de Coubertin 69627 Villeurbanne Cedex, France\label{aff61}
\and
Universit\"at Bonn, Argelander-Institut f\"ur Astronomie, Auf dem H\"ugel 71, 53121 Bonn, Germany\label{aff62}
\and
INFN-Sezione di Roma, Piazzale Aldo Moro, 2 - c/o Dipartimento di Fisica, Edificio G. Marconi, 00185 Roma, Italy\label{aff63}
\and
Aix-Marseille Universit\'e, CNRS, CNES, LAM, Marseille, France\label{aff64}
\and
Dipartimento di Fisica e Astronomia "Augusto Righi" - Alma Mater Studiorum Universit\`a di Bologna, via Piero Gobetti 93/2, 40129 Bologna, Italy\label{aff65}
\and
Department of Physics, Institute for Computational Cosmology, Durham University, South Road, Durham, DH1 3LE, UK\label{aff66}
\and
Universit\'e Paris Cit\'e, CNRS, Astroparticule et Cosmologie, 75013 Paris, France\label{aff67}
\and
CNRS-UCB International Research Laboratory, Centre Pierre Bin\'etruy, IRL2007, CPB-IN2P3, Berkeley, USA\label{aff68}
\and
Institute of Physics, Laboratory of Astrophysics, Ecole Polytechnique F\'ed\'erale de Lausanne (EPFL), Observatoire de Sauverny, 1290 Versoix, Switzerland\label{aff69}
\and
Telespazio UK S.L. for European Space Agency (ESA), Camino bajo del Castillo, s/n, Urbanizacion Villafranca del Castillo, Villanueva de la Ca\~nada, 28692 Madrid, Spain\label{aff70}
\and
European Space Agency/ESTEC, Keplerlaan 1, 2201 AZ Noordwijk, The Netherlands\label{aff71}
\and
Institut de F\'{i}sica d'Altes Energies (IFAE), The Barcelona Institute of Science and Technology, Campus UAB, 08193 Bellaterra (Barcelona), Spain\label{aff72}
\and
DARK, Niels Bohr Institute, University of Copenhagen, Jagtvej 155, 2200 Copenhagen, Denmark\label{aff73}
\and
Waterloo Centre for Astrophysics, University of Waterloo, Waterloo, Ontario N2L 3G1, Canada\label{aff74}
\and
Department of Physics and Astronomy, University of Waterloo, Waterloo, Ontario N2L 3G1, Canada\label{aff75}
\and
Perimeter Institute for Theoretical Physics, Waterloo, Ontario N2L 2Y5, Canada\label{aff76}
\and
Space Science Data Center, Italian Space Agency, via del Politecnico snc, 00133 Roma, Italy\label{aff77}
\and
Institute of Space Science, Str. Atomistilor, nr. 409 M\u{a}gurele, Ilfov, 077125, Romania\label{aff78}
\and
Dipartimento di Fisica e Astronomia "G. Galilei", Universit\`a di Padova, Via Marzolo 8, 35131 Padova, Italy\label{aff79}
\and
INFN-Padova, Via Marzolo 8, 35131 Padova, Italy\label{aff80}
\and
Instituto de F\'isica Te\'orica UAM-CSIC, Campus de Cantoblanco, 28049 Madrid, Spain\label{aff81}
\and
Institut de Recherche en Astrophysique et Plan\'etologie (IRAP), Universit\'e de Toulouse, CNRS, UPS, CNES, 14 Av. Edouard Belin, 31400 Toulouse, France\label{aff82}
\and
Universit\'e St Joseph; Faculty of Sciences, Beirut, Lebanon\label{aff83}
\and
Departamento de F\'isica, FCFM, Universidad de Chile, Blanco Encalada 2008, Santiago, Chile\label{aff84}
\and
Cosmic Dawn Center (DAWN)\label{aff85}
\and
Niels Bohr Institute, University of Copenhagen, Jagtvej 128, 2200 Copenhagen, Denmark\label{aff86}
\and
Universidad Polit\'ecnica de Cartagena, Departamento de Electr\'onica y Tecnolog\'ia de Computadoras,  Plaza del Hospital 1, 30202 Cartagena, Spain\label{aff87}
\and
European University of Technology EUt+, European Union\label{aff88}
\and
Institute of Space Sciences (ICE, CSIC), Campus UAB, Carrer de Can Magrans, s/n, 08193 Barcelona, Spain\label{aff89}
\and
Institut d'Estudis Espacials de Catalunya (IEEC),  Edifici RDIT, Campus UPC, 08860 Castelldefels, Barcelona, Spain\label{aff90}
\and
Caltech/IPAC, 1200 E. California Blvd., Pasadena, CA 91125, USA\label{aff91}
\and
Institut d'Astrophysique de Paris, 98bis Boulevard Arago, 75014, Paris, France\label{aff92}
\and
ICL, Junia, Universit\'e Catholique de Lille, LITL, 59000 Lille, France\label{aff93}}

   \authorrunning{M.~Griggio et al.}
    \titlerunning{Internal kinematics and the convective-transition gap of NGC~6397} 
   
   \date{Received 15 April 2026 / Accepted 12 May 2026}

  \abstract{
We present a `multiple-pass' data-reduction tool designed for \Euclid, based on software developed for the \textit{Hubble} Space Telescope (HST), which improves the astrometric and photometric precision for faint sources and in crowded fields. In this work, we apply it to \Euclid Early Release Observations of the Galactic globular cluster NGC\,6397. 
By combining our new catalogue with archival HST data, separated by a time span of approximately 20 years, we were able to measure high-precision proper motions and investigate the radial variations in the energy equipartition and velocity anisotropy of the cluster.  The combination of deep and wide-field observations also allowed us to derive the present-day local mass function of NGC\,6397 and to study the radial dependence of mass segregation and binary fraction.
Finally, we report the discovery of a subtle under-density of stars in the colour-magnitude diagram of NGC\,6397 around a stellar mass of $0.35\,M_\odot$ with a more than $5\sigma$ confidence level. This feature is consistent with the \textit{Gaia} M-dwarf gap discovered in Galactic field stars, but it has never previously been observed in a globular cluster. The gap is caused by the onset of full convection in stellar interiors. We demonstrate that the properties of the gap provide tight constraints on the distance to NGC\,6397 and its intrinsic metallicity dispersion, offering a new benchmark for stellar evolution models.
}

   \keywords{
   globular clusters: individual: NGC\,6397 -- 
   stars: kinematics and dynamics -- 
   stars: luminosity function, mass function --
   stars: low-mass -- 
   stars: interiors --
   proper-motions
               }

   \maketitle

\section{Introduction}

Galactic globular clusters (GCs) are among the oldest stellar systems in the Universe and represent key laboratories for a wide range of astrophysical studies, serving as fossil records of the early stages of galaxy formation and the chemical enrichment history of the Milky Way \citep{2010MNRAS.404.1203F, 2012A&ARv..20...50G,2019A&A...630L...4M}. Once thought to be simple, gravitationally bound systems composed of stars born in the same initial conditions, GCs provide valuable constraints on star formation and evolution, as well as on the dynamical processes that govern the long-term evolution of stellar systems \citep[e.g.][]{1987degc.book.....S,1997A&ARv...8....1M,2006essp.book.....S,1986ASSL..122..195R,2016RAA....16..179L}.

Over the past two decades, the classical view of GCs as simple stellar populations was impacted by the revelation that virtually all GCs host multiple stellar populations, characterised by distinct chemical abundances and photometric sequences \citep[e.g.][]{2004ApJ...605L.125B,2009A&A...505..139C,2012A&ARv..20...50G,2014ApJ...780...94C}. This discovery motivated extensive observational and theoretical efforts to understand their formation and early evolutionary histories \citep[e.g.][]{2015MNRAS.454.4197R,2018ARA&A..56...83B}. 
In parallel, multi-epoch \emph{Hubble} Space Telescope (HST) observations have delivered proper-motion measurements with a precision of a few tens of $\mu$as\,yr$^{-1}$, enabling investigations of internal cluster kinematics, including velocity-dispersion profiles, velocity anisotropy, and energy equipartition in cluster cores and at intermediate radii \citep[e.g.][]{2010ApJ...710.1063V,2014ApJ...797..115B,2015ApJ...803...29W,2022ApJ...934..150L}. By combining proper motions with deep photometry, the dynamical evolution of populations within each cluster can be explored \citep[e.g.][]{2013ApJ...771L..15R,2015ApJ...810L..13B,2024A&A...691A..94D,2025ApJ...986...80G}.

These detailed dynamical studies require the highest possible photometric and astrometric precision. In this work, we focus on NGC\,6397, one of the nearest Galactic GCs, located at a distance of about 2.48\,kpc \citep{2021MNRAS.505.5957B}. Thanks to its proximity to the Sun and its relatively small mass ($\sim$\,$8.5\times10^4\,M_\odot$; \citealt{2018MNRAS.478.1520B}) and its resultant short relaxation time, NGC\,6397 has long served as a benchmark for studies of stellar evolution and dynamics \citep{2008AJ....135.2141R,2012ApJ...761...51H}. NGC\,6397 also benefits from extensive multi-epoch HST imaging, providing an excellent test case for high-precision, cross-telescope astrometry.

More recently, the cluster has become a primary target for the \Euclid mission; \citet{Libralato24} used this dataset as a benchmark for a high-precision data reduction technique, and to showcase the synergy between \Euclid and \textit{Gaia} for achieving high-precision astrometry. Furthermore, \citet{EROGalGCs} exploited the wide-field imaging capabilities of \Euclid to perform a detailed analysis of the overall morphology of the cluster, specifically highlighting \Euclid potential in the study of extended tidal tails that trace the gravitational interaction of the cluster with the Milky Way.

Advancements in data-reduction techniques have achieved significant improvements in depth and precision by exploiting the full information content of multi-exposure imaging data \citep{2008AJ....135.2114A,2017ApJ...842....6B,2018MNRAS.481.3382N,2022ApJ...934..150L}. The `multiple-pass’ photometric approach simultaneously analyses multiple exposures using accurate astrometric registration and spatially variable, effective point-spread functions (ePSFs). It is particularly powerful in the analysis of crowded stellar fields \citep{2016ApJS..222...11S,2017ApJ...842....6B,2018ApJ...861...99L,2022ApJ...934..150L,2023MNRAS.521L..39N,2024A&A...687A..94G}.
\cite{2000PASP..112.1360A} demonstrated that this approach yields better astrometric precision than image-resampling techniques, such as drizzling, which preserve photometric accuracy but inevitably introduce correlations between pixels and lead to a degradation of positional information. 

We herein present a high-precision, multiple-pass photometric and astrometric data-reduction tool, optimised for the \Euclid telescope \citep{EuclidSkyOverview}. This approach builds upon the foundational PSF-fitting routines developed for HST by \citet{2008AJ....135.2114A}, adapting them to the unique characteristics of \Euclid's Visible (VIS; \citealt{EuclidSkyVIS}) and Near-Infrared Spectrometer and Photometer (NISP; \citealt{EuclidSkyNISP}) instruments. While \Euclid was primarily designed for large-scale cosmological surveys, its technical specifications offer an unprecedented frontier for resolved stellar-population studies \citep{EROGalGCs,Libralato24,ERONearbyGals,Larsen25,Howell25,Annibali26}. 
However, extracting the full scientific potential of \Euclid in these regions requires a departure from standard pipeline products. In dense, crowded environments such as GC cores, traditional aperture or automated photometry often suffers from severe blending and systematic biases. Our multiple-pass routine addresses these challenges, enabling the level of precision required to revisit long-standing questions in GC formation and evolution.

This paper is organised as follows. Section\,2 describes the observations used in this work. Section\,3 presents the data reduction routines and the astro-photometric catalogue. Section\,4 describes the proper-motion measurements and the analysis of the internal kinematics. Sections\,5 and 6 focus on the mass function, mass segregation and binary fraction as a function of distance from the cluster centre. Section\,7 reports the detection of a subtle feature in the low-mass colour--magnitude diagram (CMD), which we identified as the `\textit{Gaia} gap'. We discuss its physical origin and its potential as a distance and metallicity indicator. Finally, we present our conclusions in Sect.\,8.

\section{Observations}

The observations of NGC\,6397 used in this work were collected in September 2023 as part of the \Euclid Early Release Observations (ERO) Programme \citep{EROcite,EROData}, carried out with VIS and NISP. The VIS imager is a 6$\,\times\,$6 mosaic of 4k$\,\times\,$4k charge-coupled devices (CCDs) with a pixel resolution of $0{\farcs}1$ pixel$^{-1}$ and an instantaneous field of view of 0.54\,deg$^2$. This camera observes through a single broadband filter ($\IE$; 550--900 nm). Each CCD is composed of four quadrants (readouts), which we analysed as individual detectors. The NISP instrument offers both spectroscopic and photometric capabilities and is composed of 16 near-infrared 2k$\,\times\,$2k detectors, which provide a total field of view of approximately 0.57\,deg$^2$ at a $0\farcs3$ pixel$^{-1}$ scale. It covers the wavelength range 950 to 2021\,nm via three filters: $\YE$, $\JE$, and $\HE$ \citep{Scaramella-EP1}.

The data set includes: four deep (560\,s) and four short (89.5\,s) dithered exposures with VIS; and four dithered short (87.2\,s) exposures with NISP in each of the $\YE$, $\JE$, and $\HE$ filters. The data were processed with the ERO pipeline (described in detail by \citealt{EROData}) to correct instrumental effects in the raw images (mainly bias, dark, flat field, and linearity). We omitted the cosmic-ray correction of the VIS data performed by the \texttt{deepCR} code \citep{2020ApJ...889...24Z}, since empirical testing showed that this choice leads to a small but sizeable improvement in astrometry and photometry \citep[see Sect.\,2 of][]{Libralato24}.

\section{An improved data-reduction strategy: Multiple-pass photometry}
\label{sec:2pass}

\begin{figure*}[t]
    \centering
    \includegraphics[width=\textwidth]{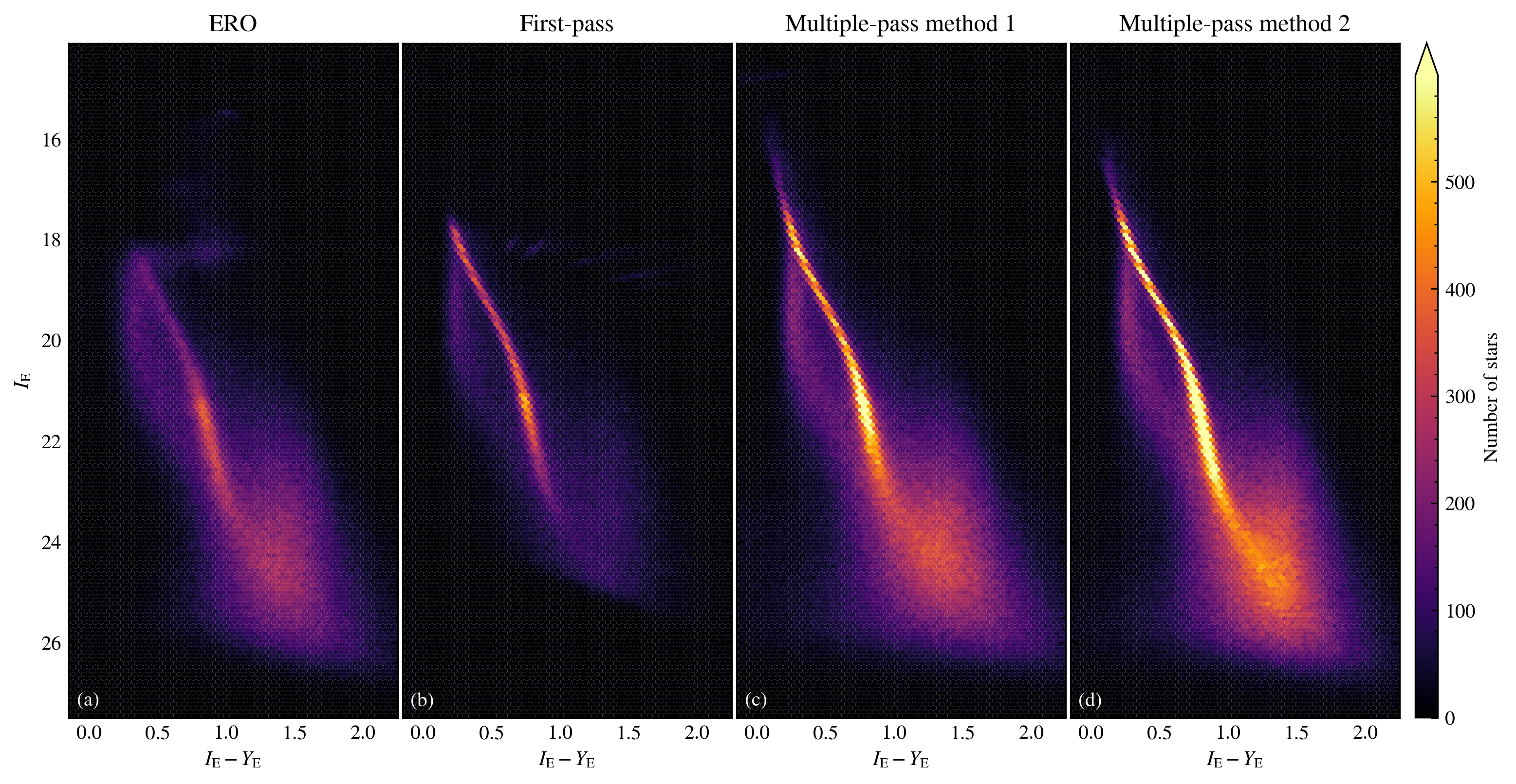}
    \caption{
    Source density across the NGC\,6397 CMD in $\IE - \YE$ filters
    obtained for the sources observed within the \Euclid field of view. Panels (a) and (b) show the CMDs obtained by \cite{EROGalGCs} and \cite{Libralato24}, respectively, from the same data. Panels (c) and (d) show the CMDs derived in this work with the two photometric methods described in Sect.\,\ref{sec:2pass}. No quality selections have been applied to these catalogues.}
    \label{fig:cmd_comp}
\end{figure*}

\begin{figure*}[t]
    \centering
    \includegraphics[width=.95\textwidth]{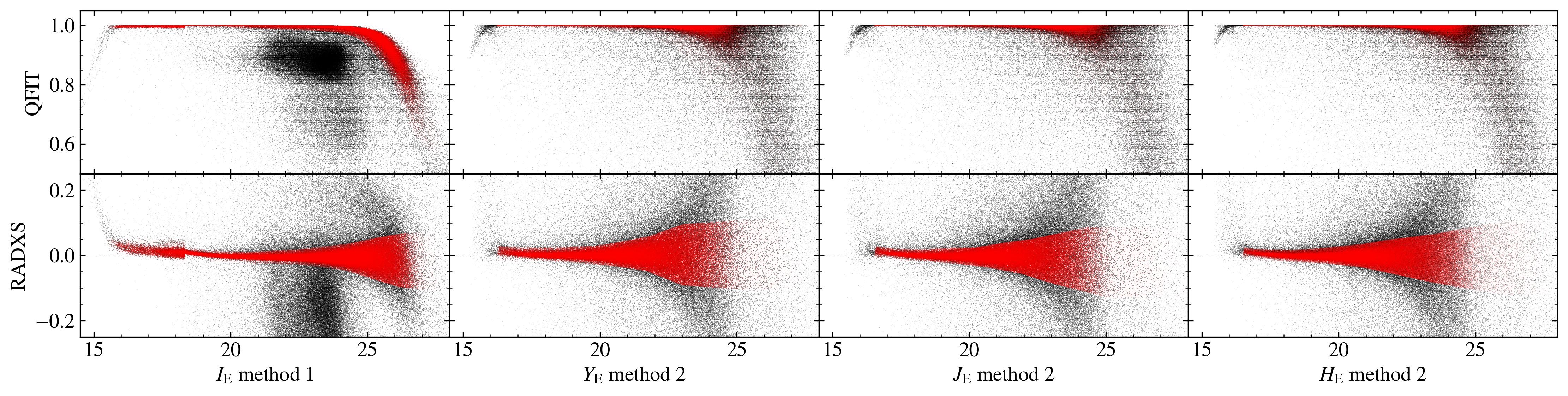}
    \caption{Quality cuts applied on the \texttt{QFIT} (\textit{top}) and \texttt{RADXS} (\textit{bottom}) parameters for all filters. Red sources are those that passed the quality cuts, while black sources are rejected according to at least one criterion. The discontinuity at $\IE$\,$\approx$\,18 corresponds to the transition point where the VIS short-exposure data were included.}
    \label{fig:ph_sel}
\end{figure*}

Our data reduction was comprised of a combination of first- and multiple-pass photometric stages \citep{2008AJ....135.2114A}, analogously to what is typically done for the reduction of HST and \emph{James Webb} Space Telescope (JWST) data \citep[e.g.][]{2017ApJ...842....6B,2018MNRAS.481.3382N,2024A&A...690A.371L}. What we refer to as `first-pass photometry' for \Euclid is detailed in \citet{Libralato24}. In brief, we obtained accurate effective point spread function (ePSF) models and geometric-distortion corrections for the VIS and NISP detectors. These products were then used to measure a preliminary set of positions and fluxes for the brightest and most isolated sources in each \Euclid image via ePSF fitting. Bright, unsaturated stars in these single-image catalogues were cross-matched with those in \textit{Gaia} \citep{2016A&A...595A...1G} Data Release 3 \citep[DR3,][]{2023A&A...674A...1G} to set up a common reference system used to combine all the catalogues \citep[see][for a detailed description of the reference-frame setup]{Libralato24}. Starting from these ePSF models, geometric-distortion corrections and reference-frame system, we proceeded with the multiple-pass stage.

The multiple-pass photometry was performed with a version of \texttt{KS2} that we tailored to the \Euclid data. The \texttt{KS2} software was originally designed for HST \citep[see][]{2017ApJ...842....6B} and comprises an evolution of the code presented in \citet{2008AJ....135.2114A}, which was later adapted to both other space telescopes \citep[e.g. JWST;][]{2023MNRAS.521L..39N} and ground-based wide-field imagers \citep{2022MNRAS.515.1841G,2023MNRAS.524..108G,2024A&A...687A..94G}.

The core design principle of \texttt{KS2} is to perform high-precision photometry by processing all available images of a field simultaneously, which is crucial for maximizing the signal-to-noise ratio (S/N) and reliably measuring faint sources that cannot be seen in individual frames. \texttt{KS2} takes as its input the transformations required to accurately map all individual exposures onto the master frame, along with an input list of bright stars (obtained from the first-pass photometry): masks around these stars mitigate artefacts such as diffraction spikes and saturated halos.

When dealing with the wide field-of-view of \Euclid, it is necessary to account for projection effects that are due to the different tangent planes of each dithered exposure \citep{2015MNRAS.450.1664L,2023MNRAS.524..108G,2024A&A...687A..94G}. Unlike narrow-field imaging, where a simple linear shift on a tangent plane might suffice, wide-field observations require a more complex geometric transformation to correct for the inherent distortion introduced when mapping the curved celestial sphere onto a flat detector plane. Therefore, the transformation step in \texttt{KS2} was upgraded to include these computations: each individual image is first projected onto the celestial sphere then subsequently de-projected onto a common tangent plane. This correction ensures that all stellar positions across the wide field of view are accurately aligned and measured consistently in the final master frame. As described in \citet{Libralato24}, the common reference frame was set as a tangent plane centred at the centre of NGC\,6397, $(\alpha,\delta) = (265{\fdg}175385,-53{\fdg}674335)$.\footnote{\url{https://people.smp.uq.edu.au/HolgerBaumgardt/globular/}}

\texttt{KS2} proceeds by dividing the observed master frame into small square tiles. By analysing simultaneously all exposures that overlap a given tile together, the software finds and measures stars, including those that are too faint to be distinctly detected in any single exposure alone. The finding process involves multiple passes over each tile, designed to iteratively improve the star list and reach fainter limits. At each iteration, the routine subtracts the flux contributions of stars measured at the previous step from the image and then searches for fainter stars in the resulting residual image. The criteria for identifying sources become progressively more relaxed from the first to the last iteration, enabling the detection of increasingly fainter objects. These identification criteria include: isolation within a certain number of pixels; significance level over the sky noise; quality of the PSF fit; and the number of coincident peaks within a given user-defined separation. The final passes are designed to detect the faintest stars possible, even those that do not generate a distinct peak in every individual exposure by using a peak-coincidence detection method \citep{2008AJ....135.2114A}.

To accurately determine the fluxes, the \Euclid version of \texttt{KS2} uses two methods. Method 1 works best when a source generates a peak (i.e. a local maximum) within its local $5\times5$ pixels, neighbour-subtracted image, and simultaneously fits position and flux with the appropriate ePSF. Method 2 is optimized for measuring the flux of very faint sources whose signal is on par with background noise in single exposures. This method uses the fixed position from the robust finding stage (wherein all the exposures are examined together to identify the faintest stars and measure their positions) and confines the flux measurement to the central four pixels of the source, expected to contain the core flux. These four pixels are weighted according to the fraction of the star's light they are expected to receive.

A particular strength of the \texttt{KS2} software, which is highly advantageous for processing \Euclid data, arises from the combined use of the VIS and NISP images. \texttt{KS2} can find stars exclusively from VIS images thanks to their higher angular resolution and depth. The stellar positions derived from VIS are then kept fixed on the NISP images during the measurement stage with method\,2. This strategy yields deeper, more reliable and more accurate near-infrared flux measurements than what could be achieved with NISP data alone.

In the final step, a robust average flux for each star is output, along with other diagnostic parameters, such as \texttt{QFIT}, \texttt{RADXS}, and flux RMS \citep{2008AJ....135.2114A}, which are essential to assess the photometric quality and nature of the measured sources. The \texttt{QFIT} parameter quantifies how well the source's light distribution matches that of the PSF. A value close to unity indicates an excellent fit, confirming the object is consistent with a clean, isolated star. Lower values might suggest problems such as blending or a profile that deviates from the expected PSF shape. \texttt{RADXS} is a shape parameter used for star-galaxy separation \citep{2008ApJ...678.1279B}. It compares the radial profile of a source to that of a PSF just outside the fitting radius. Values near zero identify a well-measured point source, while values deviating from zero indicate an extended source.

The photometry was calibrated using the first-pass catalogue of \cite{Libralato24} by cross-matching the sources and deriving a photometric zero point for each filter. The CMDs for method\,1 and 2 obtained here from \texttt{KS2} are presented in the rightmost panels of Fig.\,\ref{fig:cmd_comp}. For reference and comparison, the two CMDs on the left show the CMDs released as part of ERO catalogues \citep{EROGalGCs} and the first-pass photometry presented in \cite{Libralato24}.  The RMS of the flux provides a measure of the internal precision for the final average magnitude of a star in a given filter. The \texttt{QFIT} and \texttt{RADXS} parameters have been used to select a sample of well-measured sources in all filters, as shown in Fig.\,\ref{fig:ph_sel}. We applied a cut drawn by hand to separate the bulk of well-measured sources (in red) from poorly-measured ones (black) \citep[see e.g.][]{2017ApJ...842....6B}. The magnitudes shown in the $x$-axis are obtain from method 1 for VIS and method 2 for NISP. In the analysis, we used method-2 photometry for NISP filters since is proven to be superior to method 1 (see Sect.\,\ref{sec:io}).

\section{Internal kinematics}

\begin{figure}
    \centering
    \includegraphics[width=0.95\columnwidth]{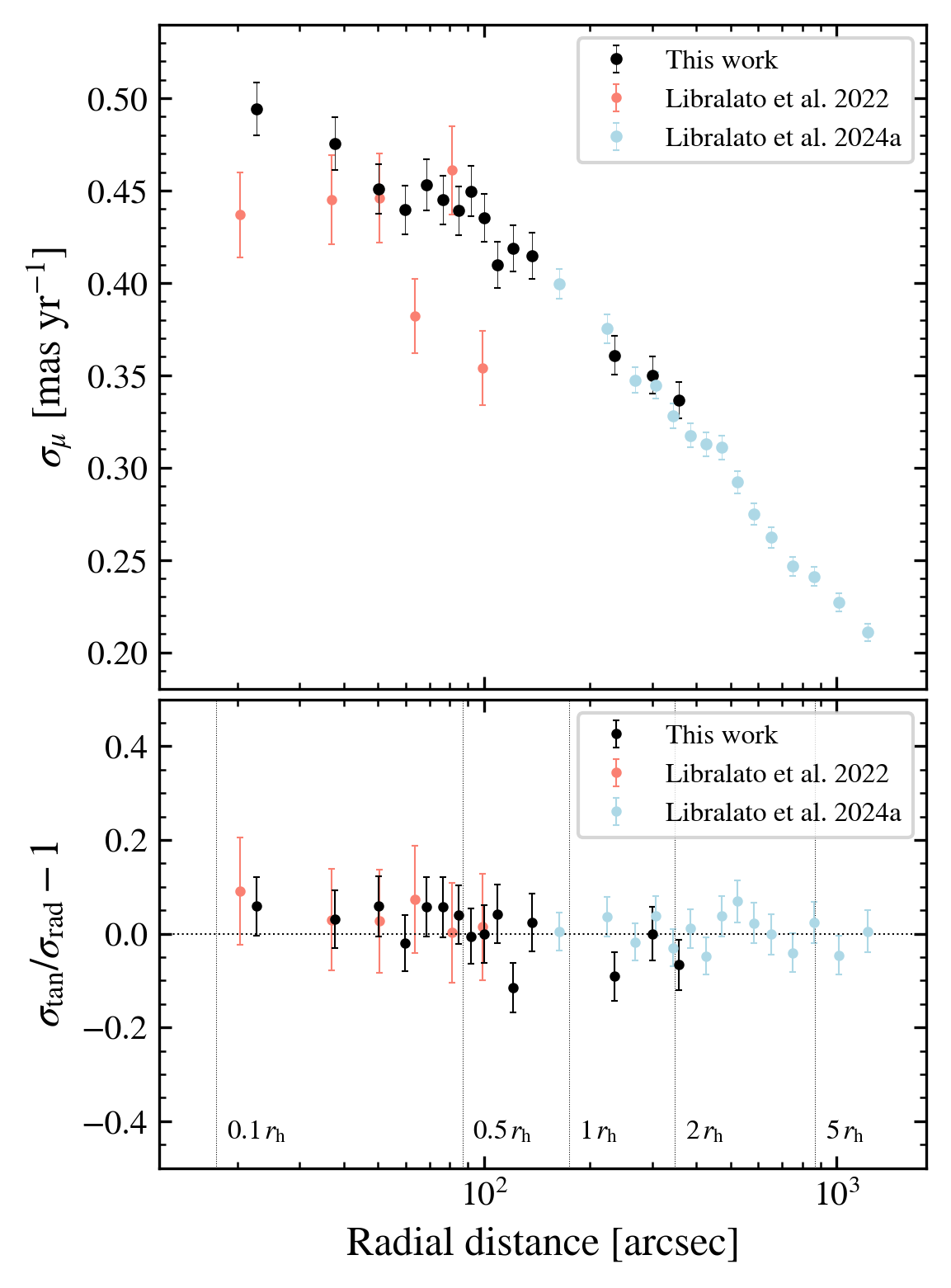}
    \caption{\emph{Top:} Combined proper-motion velocity dispersion profile for stars in NGC\,6397. Black points represent our \Euclid--HST measurements. Light red and light blue points are from \cite{2022ApJ...934..150L} and \cite{Libralato24}, respectively.
    \emph{Bottom:} Velocity anisotropy profile for the data points.}
    \label{fig:vdisp}
\end{figure}

\begin{figure}
    \centering
    \includegraphics[width=0.95\columnwidth]{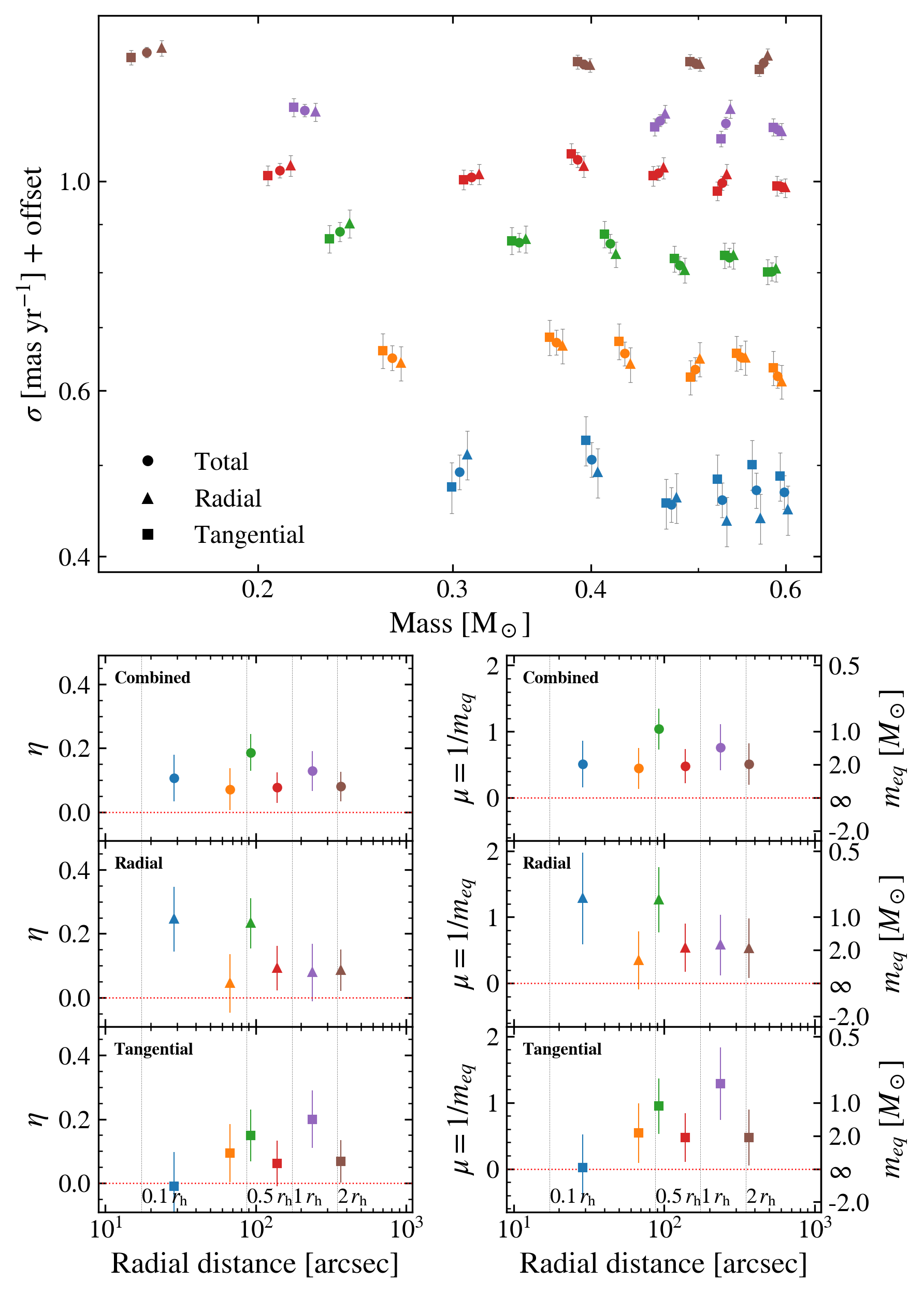}
    \caption{
    \emph{Top:} Combined velocity dispersion as a function of stellar mass at different radial intervals, colour-coded as the points in the lower panel, shifted by constant factors, except for the baseline blue points.
    \emph{Bottom:} Energy equipartition parameters $\eta$ (bottom-left panels) and $\mu=1/m_{\rm eq}$ (bottom-right panels) as a function of radial distance from the centre of the cluster.}
    \label{fig:equip}
\end{figure}

To study the internal kinematics of NGC\,6397, we coupled our catalogues with HST archival data of two fields, one centred on the core of the cluster, and one at about $3^\prime$ from the core. The HST data were reduced as described in \cite{2017ApJ...842....6B} and \cite{2022ApJ...934..150L}, using a combination of first- and multiple-pass photometry, as also done in this paper for the \Euclid data. We then cross-identified the same sources between the \Euclid and HST data.\looseness=-1

We derived proper motions from the combined HST--\Euclid data set following the iterative method introduced in \cite{2014ApJ...797..115B} and applied in several other works \citep[e.g.][]{2022ApJ...934..150L,2025ApJ...986...80G}. We treated each exposure as an independent epoch and transformed the corresponding distortion-corrected positions into a common reference frame using linear transformations. These transformations are derived by cross-identifying stars between the single-image catalogues and the master catalogue, while using only a local network of cluster members. The transformed positions on the master frame at different epochs were fitted with a linear relation (with each coordinate fitted independently). The fitted slope is an estimate of the motion, while the intercept gives the position at our chosen reference epoch 2016.0. 

We refined the master frame and the resulting proper motions through an iterative cycle. In each step, we used the fitted motions to propagate stellar positions to the mean epoch of each exposure group, thereby creating an improved reference catalogue, which accounts for the motion of each source. The linear transformations are then recalculated, and the process is repeated until the positional residuals in the master catalogue positions between successive iterations become less than 0.01 pixels. By only using cluster members as reference sources for these transformations, our derived motions are relative to the bulk motion of the cluster, and any systemic rotation or global contraction and expansion shared by the cluster members is absorbed into the transformation, leaving only the internal relative dispersions.

To account for systematic errors, we applied a multi-step correction process. First, we corrected for low-frequency temporal effects by grouping the data according to their temporal baselines. For each group, we calculated the median motion of bona-fide cluster members (identified as within 1.25\,mas\,yr$^{-1}$ from the bulk motion of the cluster) and subtracted this value to ensure a common zero-point. We then addressed high-frequency, spatially- and magnitude-dependent systematics primarily arising from charge transfer efficiency degradation in HST data and residual geometric distortion. We applied a local correction for each star by calculating the average motion of its 200 (in the central field) or 100 (in the external field) nearest-neighbour cluster members within a 1250-pixel radius and $\pm$\,1 magnitude. Finally, we applied a quality selection on the catalogue. We retained only stars with a measurement rejection rate under 25\% and a reduced $\chi^2<4$ in both coordinates. Following the procedure in \cite{2014ApJ...797..115B}, we iteratively rejected sources with proper-motion uncertainties greater than 50\% of the local velocity dispersion to produce the final high-precision kinematic measurements.\looseness=-1

The combined velocity dispersion profile of the cluster is shown in Fig.\,\ref{fig:vdisp} (top panel). The masses of the stars used in the analysis span the range 0.1--0.6\,$M_\odot$.
We divided the data into equally-populated radial bins, with about 300 stars per bin. Black points represent our \Euclid-HST measurements. The light red points are from \cite{2022ApJ...934..150L}, are based on HST data alone, and were calculated for stars at the mainsequence turn-off or brighter. Light-blue points are from \cite{Libralato24} and were derived by combining \Euclid and \textit{Gaia} data. Since they were limited by \textit{Gaia}, the covered mass range is $\sim$\,0.46--0.58\,$M_\odot$. 
The bottom panel of Fig.\,\ref{fig:vdisp} shows the velocity anisotropy profile, with the colour coding as in the top panel. The data show no clear indication of anisotropy across the explored radial range.

\subsection{Energy equipartition}
To study the radial variation of energy equipartition, we split the data into distinct radial bins. Within each radial bin, sources were further divided into equally populated magnitude bins to ensure statistical consistency. Specifically, we divided the central field into four radial bins, each containing six magnitude bins of approximately 140 stars. For the external region, we used two radial bins, each subdivided into four magnitude bins of approximately 115 stars. In Fig.\,\ref{fig:equip}, we present the radial variation of the degree of energy equipartition. The top panel displays the dependence of velocity dispersion on stellar mass for individual radial bins (offset vertically for clarity).

Individual stellar masses were estimated using an $\alpha$-enhanced ($[\alpha/\text{Fe}] = +0.4$) BaSTI-IAC isochrone \citep{2018ApJ...856..125H,2021ApJ...908..102P} with an age of $12.6\,\text{Gyr}$ and metallicity $[\text{Fe/H}] = -1.9$, assuming a distance of 2.45\,kpc and $E(B-V)=0.18$ \citep[compatible with values found in literature, e.g,][]{2015ApJ...812..149W,matteo_NGC6397,2018MNRAS.478.1520B,2020MNRAS.493.3363H,2023MNRAS.526.5628G,EROGalGCs}.

In the bottom panels, we show the radial variation of equipartition across the cluster for the combined (circles), radial (triangles), and tangential (squares) components. The left column shows the parameter $\eta$, defined from a power-law fit of the relation $\sigma \propto m^{-\eta}$ 
\citep[see e.g.][]{2013MNRAS.435.3272T}. The right column shows the equipartition mass parameter $m_{\rm eq}$ introduced by \cite{2016MNRAS.458.3644B} and its reciprocal $\mu = 1/m_{\rm eq}$, which allows for a continuous estimation of the degree of equipartition, accommodating both positive and negative slopes in the mass–velocity-dispersion relation \citep[see][]{2023MNRAS.525.3136A}. The relation between the velocity dispersion and $m_{\rm eq}$ is given by
\begin{equation*}
    \sigma(m) = \begin{cases}
        \sigma_0 \exp{\left(-\frac{1}{2} \frac{m}{m_{\rm eq}}\right)} \quad {\rm if} \,\, m\leq m_{\rm eq}\\
        \sigma_0 \exp{\left(-\frac{1}{2}\right)} \left(\frac{m}{m_{\rm eq}}\right)^{-\frac{1}{2}} \quad {\rm if} \,\, m> m_{\rm eq} \,,
    \end{cases}
\end{equation*}
where $\sigma_0$ is the velocity dispersion at $m=0$. For stellar masses larger than $m_{\rm eq}$ stars are in full energy equipartition, while for stellar masses smaller than $m_{\rm eq}$ stars have not reached full equipartition and the velocity dispersion depends on mass according to the exponential function reported above. Smaller values of $m_{\rm eq}$ indicate  a more advanced stage of a system's evolution towards complete energy equipartition.

Our analysis reveals a degree of energy equipartition generally consistent with the results of \cite{2022ApJ...936..154W}.
Overall, the degree of energy equipartition does not vary with distance from the centre of the cluster. While some mild radial decrease of the degree of equipartition is expected during some phases of cluster evolution \citep[see][]{2013MNRAS.435.3272T,2023MNRAS.525.3136A}, clusters in the advanced stages of their evolution may be characterised by similar values of $\eta$ at all radial distances \citep{2013MNRAS.435.3272T}. NGC\,6397 is a dynamically old system and may indeed have reached a state characterised by no radial variation of $\eta$ (although data from \citealt{2022ApJ...936..154W} hint at a possible radial variation); data spanning a broader radial range than that available here are necessary for a more comprehensive study of this issue.
Finally we have calculated the degree of energy equipartition using, separately, the radial and the tangential components of the velocity dispersion. A number of theoretical studies \citep{2021MNRAS.504L..12P,2022MNRAS.509.3815P,2024A&A...689A.313P,2024MNRAS.534.2397L} have shown that the degree of energy equipartition in the radial and tangential velocity dispersion may differ, particularly in a cluster's early and intermediate evolutionary phases (see \citealt{2025ApJ...986...80G} and \citealt{2025A&A...698A.209Z} for observational studies of finding hints of these differences in NGC\,2808 and 47\,Tuc, respectively). The results of our analysis are shown in the bottom two columns of Fig.\,\ref{fig:equip} and do not reveal any significant difference between  the radial and tangential degree of equipartition. This outcome is in general agreement with what is expected for dynamically old systems such as NGC\,6397.

\subsection{Kinematic concentration}

\cite{2018MNRAS.475L..96B} introduced the kinematic concentration parameter $c_k$, defined as
\begin{equation*}
    c_k = \frac{m_{\rm eq}(r<r_{50})}{m_{\rm eq}(r_{50})}\,,
\end{equation*}
where $m_{\rm eq}(r<r_{50})$ is the global equipartition mass defined using all stars inside the half-mass radius and $m_{\rm eq}(r_{50})$ is the value of the equipartition mass defined in a shell around the half-mass radius. \cite{2018MNRAS.475L..96B} showed that this parameter provides a diagnostic for core-collapse, based entirely on the internal kinematics of a cluster: a value of $c_k$ greater than one indicates that the cluster has reached the core-collapsed state.

As in \cite{2018ApJ...861...99L}, we adopted the half-light radius $r_{\rm h}$ instead of the 50\% Lagrangian radius because it is directly observable. Since our HST data have a gap between 0.75\,$r_{\rm h}$ and $r_{\rm h}$, we cannot estimate the value of $c_k$ following the exact definition proposed by \cite{2018MNRAS.475L..96B}. Instead, we calculated $c_k$ by estimating $m_{\rm eq}(r_{50})$ using stars between $r_{\rm h}$ and 1.4\,$r_{\rm h}$, and $m_{\rm eq}(r<r_{50})$ using stars at $r<0.75\,r_{\rm h}$, obtaining $c_k = 1.1 \pm 0.7$. We also repeated the estimate of $c_k$ with a different interval, $0.5\,r_{\rm h}< r < 1.5\,r_{\rm h}$ for $m_{\rm eq}(r_{50})$ and $r<0.5\,r_{\rm h}$ for $m_{\rm eq}(r<r_{50})$. The resulting kinematic concentration parameter is $c_k = 1.6 \pm 0.7$. Both these values are consistent with the core-collapsed state of NGC\,6397 \citep{1995ApJ...439..695C,2009MNRAS.397L..46H,2022MNRAS.514..806V}.

\section{The present-day local mass function}
\label{sec:mf}
\begin{figure}
    \centering
    \includegraphics[width=\columnwidth]{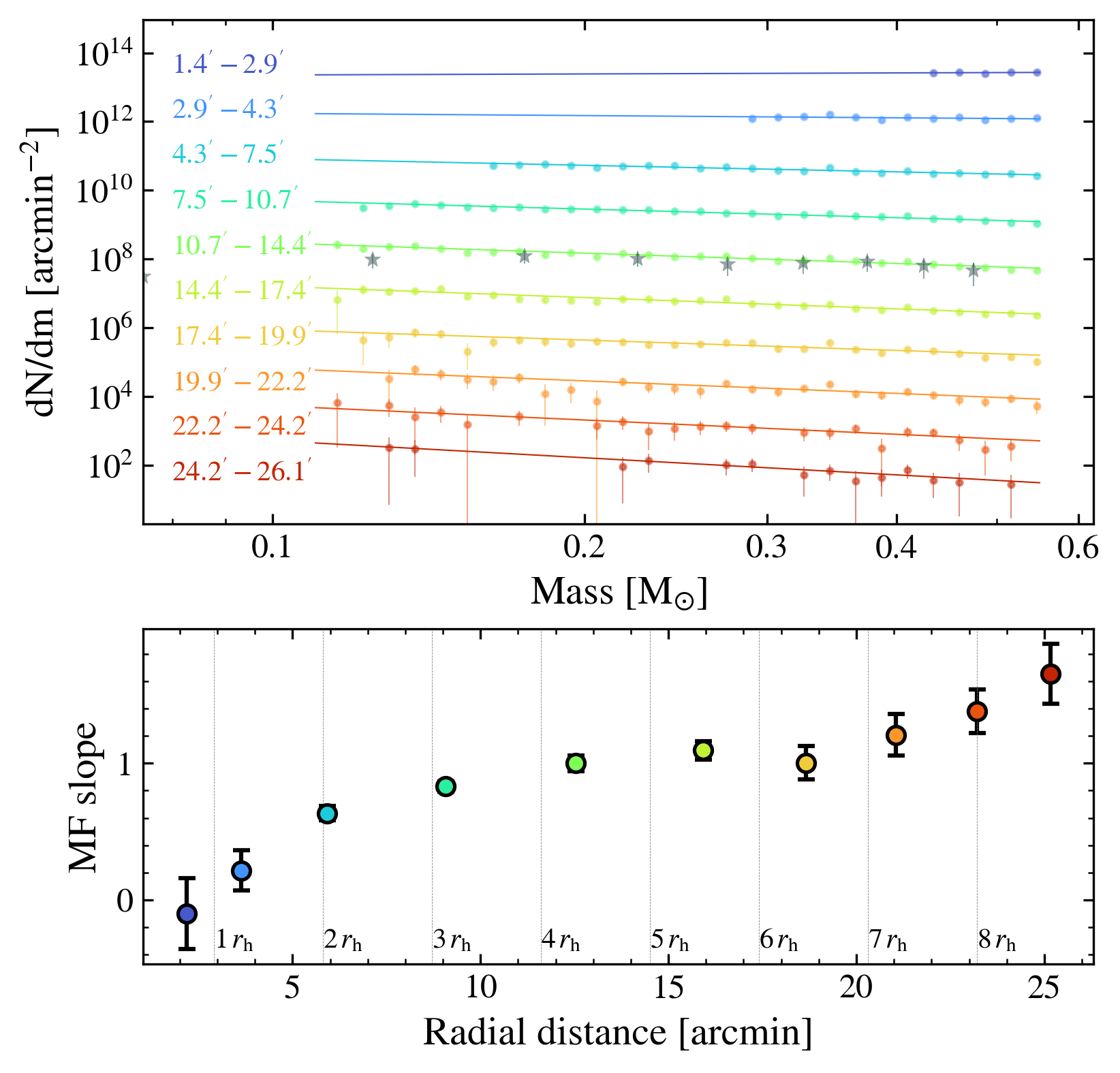}
    \caption{\emph{Top:} Decontaminated and completeness-corrected stellar mass functions for different radial intervals, shifted for clarity. Points with error bars represent the observed data with $1\,\sigma$ uncertainties. Star markers show data from \cite{2024A&A...690A.371L} for comparison, shifted to the corresponding radial interval.
    \emph{Bottom:} Mass-function slope as a function of radial distance from the cluster centre.}
    \label{fig:mf}
\end{figure}

To derive the cluster's present-day local mass function (MF), we split the data into concentric annuli extending from $0.5 \, r_{\rm h}$ to $9 \, r_{\rm h}$. To isolate the cluster sequence, we implemented a CMD selection using a fiducial envelope (defined by eye) that encompasses the cluster main sequence in the $\IE-\YE$ versus $\IE$ space. Field decontamination was performed by comparing the target annuli to a designated control field located between $9$ and $12 \, r_{\rm h}$. We calculated the decontaminated number density by subtracting from the observed density the number of stars measured in the control region, normalised by the area of each annulus. Note that the control field we used is still inside the tidal radius of the cluster, and thus our MF could be slightly underestimated. 

We used a spatially dependent completeness correction to account for variations in crowding between the inner region of the cluster and its outskirts. We used 2D completeness maps for the \IE and \YE filters stacked into 3D grids (position and magnitude), where a trilinear interpolation determined the detection probability for each star based on its position and magnitude. The total completeness also accounts for the quality selection cuts described in Sect.\,\ref{sec:2pass}.
The completeness factor for each mass bin was determined by sampling 10\,000 points uniformly distributed within each annulus to generate a median completeness value, allowing us to obtain a robust estimate of the completeness in each region. 
To ensure data reliability, we excluded all bins where the completeness factor fell below $40\%$. 

The resulting MFs (Fig.\,\ref{fig:mf}) were fitted with power-law indexes $\alpha$ ($\mathrm{d}N/\mathrm{d}m \propto m^{-\alpha}$) over the available mass range using a weighted linear least-squares fit. The bottom panel of Fig.\,\ref{fig:mf} reveals a clear signature of mass segregation characterised by a significant flattening of the MF in the core and a progressive steepening in the external regions (see also \citealt{1995ApJ...452L..33K} for the first evidence of a radial variation of the mass function).

To estimate the extent of mass loss due to two-body relaxation and the effects of the external tidal field on the MF, it is necessary to calculate the slope of the global present-day MF (PDMF) and compare it with that of the initial MF \citep[IMF; see e.g.][]{1997MNRAS.289..898V,2003MNRAS.340..227B}. If we estimate the slope of the global PDMF simply by combining all the stars used to measure the local MF shown in Fig.\,\ref{fig:mf} we find a slope equal to 0.6. However, this simple procedure leads to a biased estimate of the MF slope since it does not include low-mass stars in the cluster's inner regions (see the inner local MFs in the top panel of Fig.\,\ref{fig:mf}). If we estimate the contribution of low-mass stars in the inner regions by extrapolating the MF fit (shown in the top panel of Fig.\,\ref{fig:mf}) to low masses, we obtain a global PDMF slope equal to 0.81 (although this estimate still does not include all the stars in the innermost regions, $r<1\arcmin\!\!.4$).

This slope is substantially shallower than slope of 1.3 given by a \cite{2001MNRAS.322..231K} IMF, indicating that the MF of NGC\,6397 has been significantly flattened by the preferential loss of low-mass stars due to the effects of two-body relaxation and the Galactic tidal field, and that the cluster must have lost a large fraction of its initial mass \citep[see e.g.][]{1997MNRAS.289..898V,2003MNRAS.340..227B}. Indeed, the estimated initial mass of NGC\,6397 is about six times larger than its present-day mass according to the models of \cite{2009MNRAS.395.1173G} and about
3.5 times larger than the present-day mass according to the study
of \cite{2025MNRAS.537.1807A}.

\section{The binary fraction}

The fraction of binaries is an essential component of the
formation and evolution of any stellar system. In particular,
in dynamically active aggregates, such as GCs, binaries are
thought to promote the formation of exotic objects such as blue straggler stars,
X-ray sources, and millisecond pulsars (see e.g. \citealt{mcrea64,paresce92,pooley06,heinke03,ferraro09}). Binaries are typically more massive than the average single star in GCs ($\left<m\right> \approx 0.3$~$M_\odot$), and therefore they tend to sink toward the centre of GCs because of dynamical friction. As a consequence, their radial distribution can also be a useful tool to constrain the dynamical state of GCs. While the binary fraction in the innermost regions of GCs has been more extensively investigated mainly thanks to HST data (e.g. \citealt{sollima07,milone12}), little was known until very recently \citep{Cadelano2026} about the radial variation of the binary fraction extending to the clusters' outskirts \citep{dalessandro11,dalessandro15,beccari13,cordoni25}.

The central binary fraction of NGC~6397 was derived by \citet{milone12} by using deep HST data. Here we take advantage of the deep and wide field \Euclid images to compute the binary fraction in a complementary region out to a cluster-centric distance $r=1500\arcsec$. To this aim, we used the so-called `secondary MS' approach (e.g. \citealt{romani91,bellazzini02}).
The basic idea is that the magnitude of the binary system corresponds to the luminosity of the primary star (more massive) increased by that of the companion. Stars on the MS obey a mass–luminosity relation, hence the luminosity of the binary system is a function of the mass ratio $q=m_2/m_1$ of the two components (where $m_1$ and $m_2$ are the masses of the primary and secondary, respectively). Since $q$ can assume any value between zero and unity, binaries broaden the single-star MS in CMDs at higher luminosities.

\begin{figure}[t]
    \centering
    \includegraphics[width=1\columnwidth]{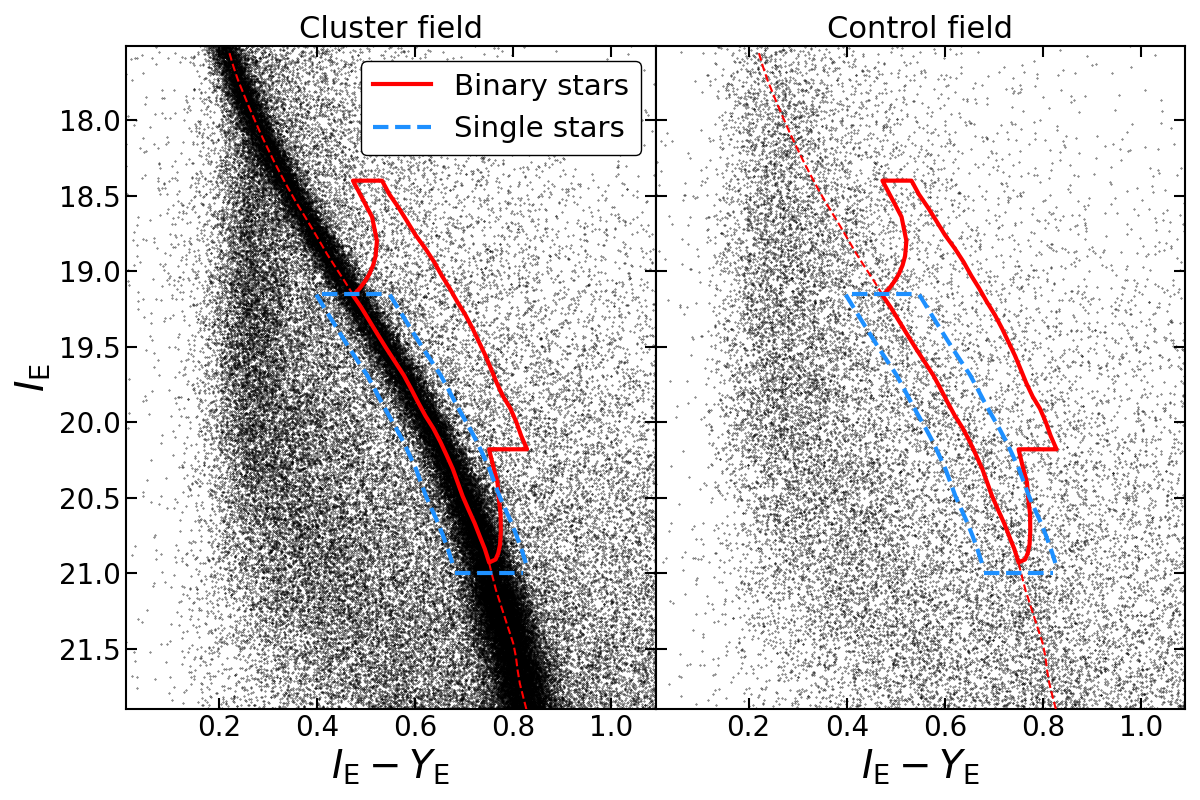}
    \caption{Zoomed-in view of the cluster MS with the selection regions for single and binary stars highlighted. The left panel panel shows the bulk cluster population and the right panel shows the control field.}
    \label{fig:cmd_bin}
\end{figure}

We estimated the minimum binary fraction $\xi$, which is the fraction of binaries with a mass ratio $q_{\rm min}$ large enough to make them clearly distinguishable from single MS stars following the approach described in \citet[][see also \citealt{dalessandro11,dalessandro15}]{bellazzini02}. We refer the reader to those papers for details.
The analysis was performed in the (\IE--\YE,\,\IE) CMD and for stars with $19<\IE<21$, where the completeness $C$ is larger than $50\%$ and the MS is not vertical. This interval corresponds to a single-star mass range of about 0.3--0.5\,$M_\odot$ (see Fig.~\ref{fig:cmd_bin}).\looseness=-1

For the derivation of the minimum binary fraction, we considered binaries located at three times the photometric error from the MS ridge line, which corresponds to $q_{\rm min} \approx 0.5$ in the adopted magnitude range. 
Contamination from Galactic field interlopers was derived by using a control field located at $r>9\,r_{\rm h}$ from the cluster centre. 
While NGC~6397 extends more than the field of view covered by the adopted \Euclid observations \citep[see][]{EROGalGCs},  the right panel of Fig.~\ref{fig:cmd_bin} shows there is only a marginal 
cluster contribution to the CMD in the control field. The contamination from blended sources has been accounted for by means of artificial-star tests (see Sect.~\ref{sec:artificial}).

\begin{figure}[t]
    \centering
    \includegraphics[width=\columnwidth]{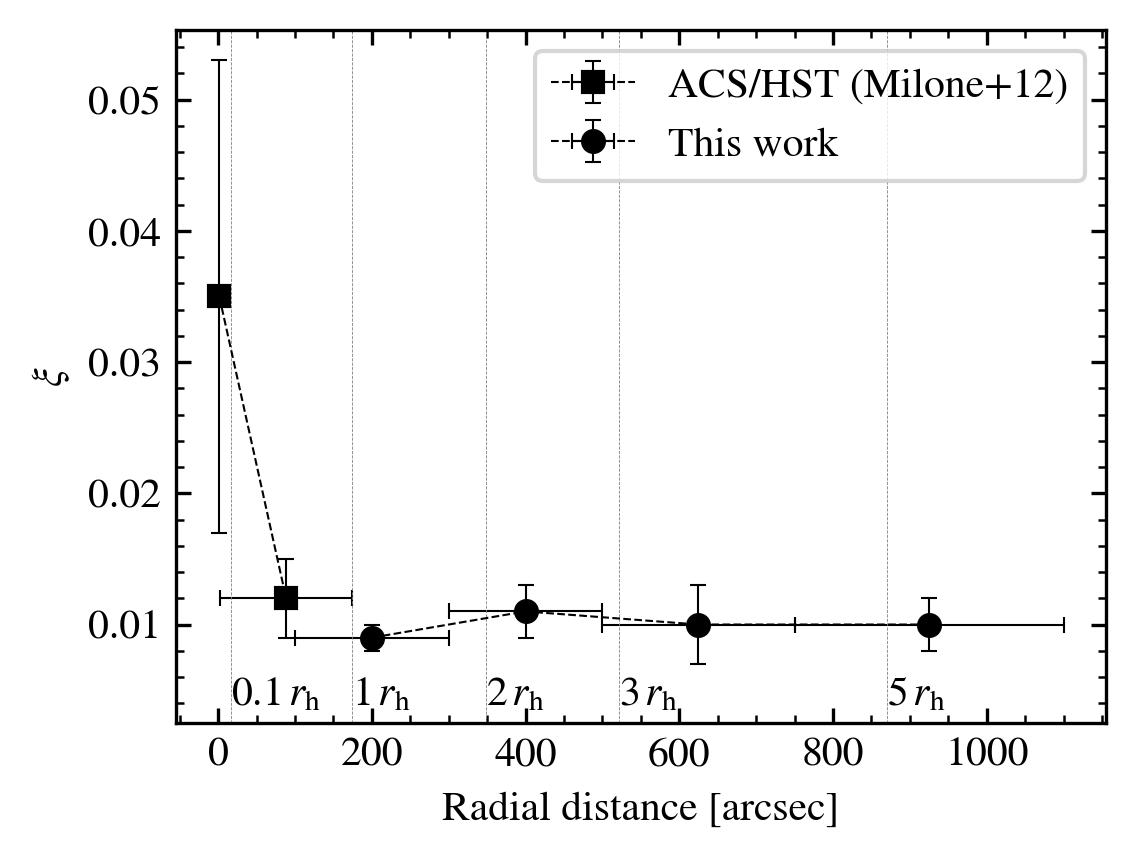}
    \caption{Minimum binary fraction radial distribution. Squares correspond to binary fractions obtained with HST by \citet{milone12}.}
    \label{fig:bin_rad}
\end{figure}

Figure \ref{fig:bin_rad} shows the radial distributions of the derived binary fraction.  For the innermost regions we adopted the binary fraction of \citet{milone12}. Consistently with very recent results presented by \cite{Cadelano2026}, the binary fraction radial distribution of NGC~6397 shows no sign of bimodality. It shows a peak in the central regions ($\xi_{\rm min} \approx 4\%$),  then 
a monotonically decreasing trend out to $R \approx 1300\arcsec$ where $\xi \approx 1\%$.
As discussed by \cite{Cadelano2026} and \citep{2026arXiv260207119B}, the binary fraction radial distributions are the result of the combined effect of cluster dynamical evolution and mass segregation, binary disruption, and of the presence of multiple populations with different spatial distributions (e.g. \citealt{2013ApJ...771L..15R,2015ApJ...810L..13B,2024A&A...691A..94D,2025ApJ...986...80G}).
Interestingly in this context, NGC\,6397 is a post-core-collapse cluster. In agreement with the observations (Fig.~\ref{fig:bin_rad}), in such dynamically old systems the binary fraction radial distribution is expected to monotonically decrease at increasing radial distances. Hence, the behaviour of the observed binary radial distribution provides additional and independent clues about the advanced dynamical stage of NGC~6397.

\section{A gap in the lower main sequence}

\begin{figure}
    \centering
    \includegraphics[width=\columnwidth]{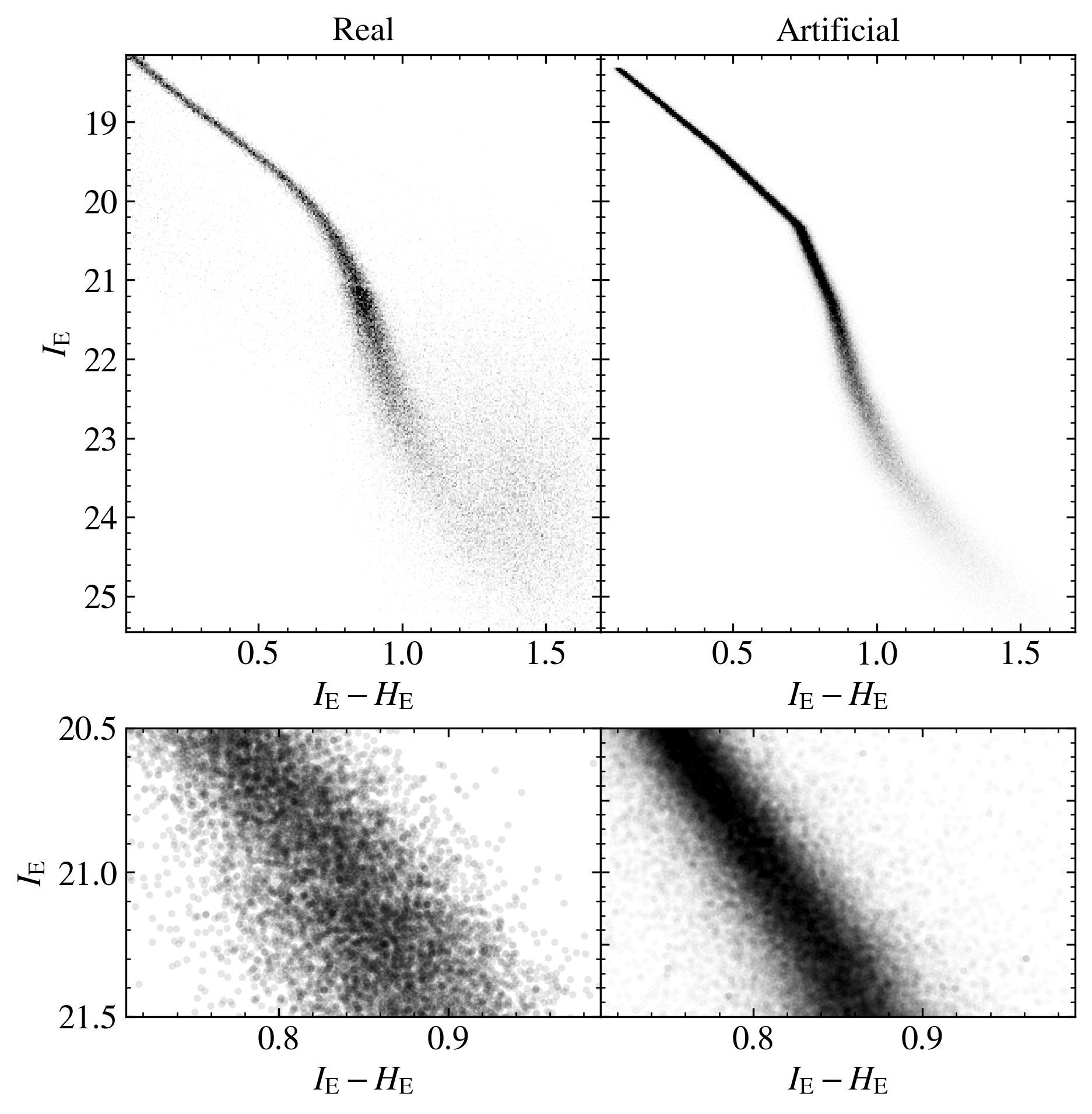}
    \caption{\textit{Top}: Hess diagram of the $\IE$ versus $\IE - \HE$ CMD for real stars (left) and artificial stars (right) that passed our photometric cuts. \textit{Bottom}: CMD for proper-motion selected stars (left) and artificial stars (right), zoomed in on the location of the gap.}
    \label{fig:gap}
\end{figure}

\begin{figure}
    \centering
    \includegraphics[width=.9\columnwidth]{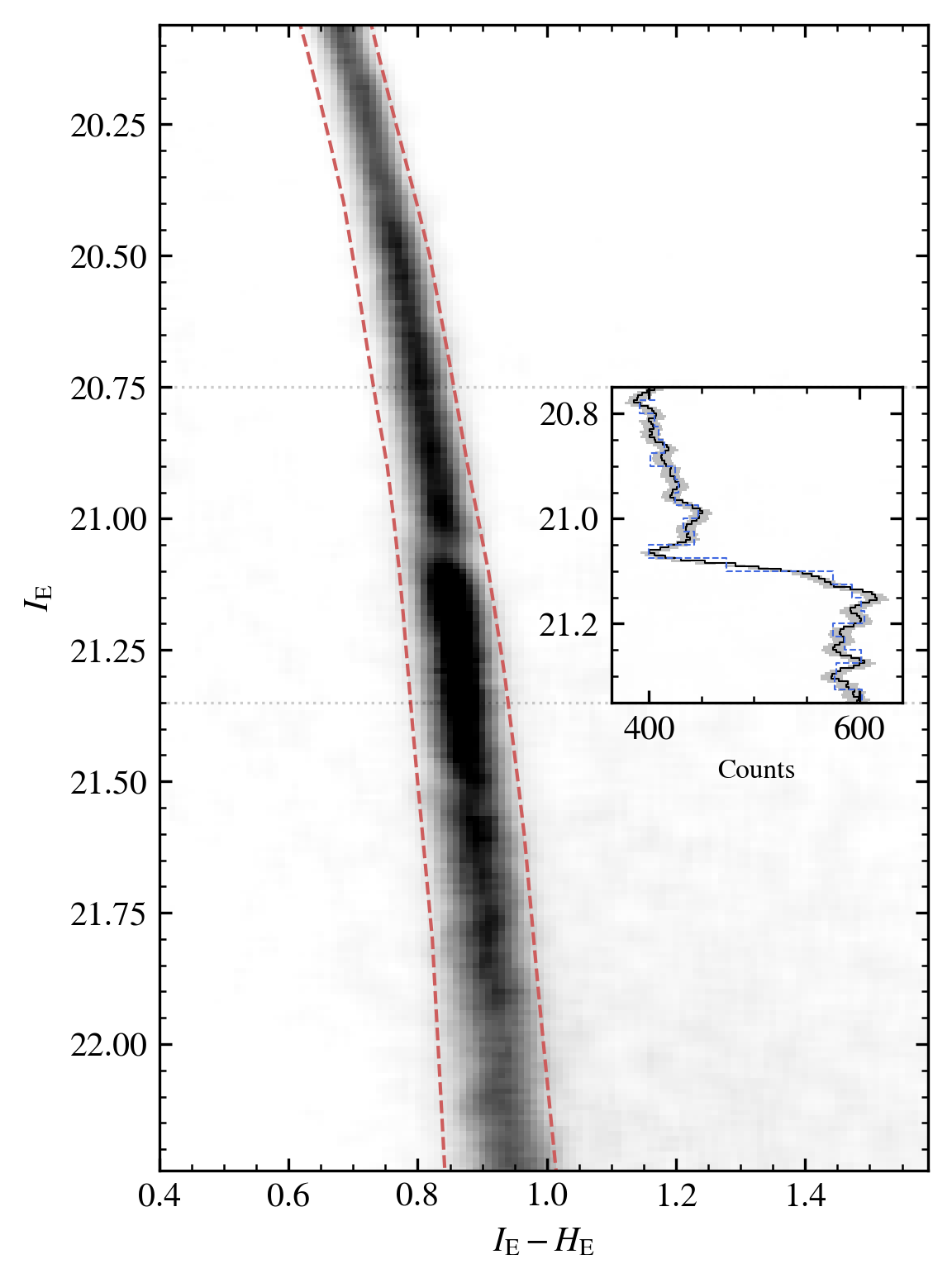}
    \caption{Greyscale of the stellar density in the $\IE$ versus $\IE - \HE$ CMD of NGC\,6397 for well measured sources. The inset panel shows the luminosity function of the region around the gap. See text for details.}
    \label{fig:gap_lf}
\end{figure}

We observe a narrow ($\Delta \IE \approx 0.05$ mag) gap (an under-density of stars in the CMD) in the cluster MS near the apparent magnitude of $\IE \approx 21.06$. We also see a `step' (discontinuous change) in star counts per unit magnitude across the gap. As we will demonstrate in Sect.\,\ref{sec:gap_physical}, this feature is closely related to comparable gap and `step' identified in the CMD of field stars with \textit{Gaia} and, at much lower significance, 2MASS photometry \citep{2018ApJ...861L..11J,Jao_step}. The gap is associated with the transition of low-mass stars from being fully convective to having a radiative zone \citep{MacDonald_gap_models,Baraffe_gap_models,2021ApJ...907...53F,2021A&A...650A.184M,2023ApJ...944..129B}, and it provides a unique window into stellar interiors on the lower main sequence.

This study is the first detection of the gap in a star cluster, uniquely enabled by the photometric precision achieved by our data reduction and the wide field of view of the \Euclid telescope. The field of view, in particular, provides a large photometric sample size, which proved crucial to this discovery. By repeating our data analysis (to be described in Sect.\,\ref{sec:gap_morphology}) on randomly drawn subsets of out dataset, it can be demonstrated that a statistically significant detection ($\geq 3\,\sigma$ confidence) of the gap in NGC\,6397 requires at least $\approx 1500$ stars with reliable photometry within $0.1$ mag of the gap, which corresponds to $65\%$ of our actual sample. It has been recently suggested \citep{2026arXiv260221882M} that the gap is entirely absent from GCs based on its non-observation in deep HST and JWST photometry; however, such outcome is fully expected given that the fields of view of these facilities are hundreds of times smaller than that of \Euclid.

The gap is shown in Fig.\,\ref{fig:gap}, which displays the so-called `Hess diagram' in $\IE - \HE$ space (top left) alongside a standard CMD zoomed-in near the gap location (bottom left). The histogram was derived using all stars that passed our photometric cuts in the $\IE$ and $\HE$ filters. In addition, the bottom left panel of Fig.\,\ref{fig:gap} has also been filtered by cluster membership. To remove field stars, we coupled the ground-based catalogue from \cite{2019MNRAS.485.3042S} with our \Euclid catalogue to derive proper motions and isolate a sample of member stars. The gap therefore cannot arise from field contamination. Furthermore, the absence of this feature in our artificial star tests (Fig.\,\ref{fig:gap}, right panels) indicates that it is not a result of localised completeness issues or instrumental systematics. 

In Fig.\,\ref{fig:gap_lf} we use a sliding square window to better highlight the gap in the $\IE$ versus $\IE - \HE$ CMD. For each point on a high-resolution grid with a sampling step of 0.01 in both colour and magnitude, the local density is calculated by counting all stars within a window of $2h_{\rm colour} \times 2h_{\rm mag}$, where $h = 0.025$ represents the window half-width. This approach is used strictly for visualisation purposes. In the inset of Fig.\,\ref{fig:gap_lf} we show the luminosity function for the region around the gap and delimited by the two fiducial lines (in red in the CMD).
The blue line in the inset shows the histogram of the $\IE$ magnitude, while the black line shows an averaged shifted histogram. We compute five independent histograms using a fixed bin width of 0.025, with each realisation offset by a sub-step of 0.005. With this method, we minimise the statistical noise and bin-edge bias associated with arbitrary bin edges, resulting in a smoothed count that better preserves the location of narrow physical features. The shaded light grey area represents the 1\,$\sigma$ errors.

While the gap can be discerned in the inset of Fig.\,\ref{fig:gap_lf}, it appears less pronounced in the 1D $\IE$ distribution than it does in the 2D CMD. This is largely because the exact magnitude of the gap varies with colour (the gap is `smeared' across a range of $\IE$), which makes it appear shallower and distorts its shape. However, the inset clearly displays a `step' in the luminosity function (i.e. a discontinuous change in star counts), which coincides with the faint edge of the gap.

\begin{figure}
    \centering
    \includegraphics[width=\columnwidth]{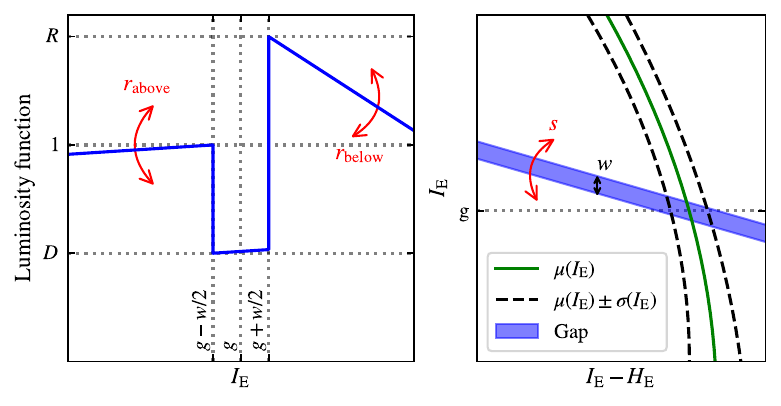}
    \caption{Schematic representation of the 13-parameter empirical gap model adopted in Sect.\,\ref{sec:gap_morphology} in luminosity (\textit{left}) and colour-magnitude (\textit{right}) spaces. The luminosity function is shown for the central colour of the gap, and is normalized to unity at the bright edge.}
    \label{fig:schematic}
\end{figure}

\subsection{Morphology of the gap} \label{sec:gap_morphology}

To infer the properties of the gap in NGC\,6397, we constructed a 13-parameter empirical likelihood model and fitted it to the observed distribution of stars in the $\IE$ versus $\IE - \HE$ CMD in the range $20<\IE<22$. In this model, the gap is approximated as a rectangular dip in the colour-magnitude diagram, whose central magnitude varies linearly with colour. This functional form is motivated by the morphology of the theoretical luminosity function derived in Sect.\,\ref{sec:gap_physical}. We adopt this empirical parameterization, rather than fitting the theoretical luminosity function directly for two reasons. First, our physical models are computed only in magnitude space for this preliminary analysis, while forward modelling of the complete colour-magnitude space is deferred to a future study. Second, the empirical model provides additional flexibility, allowing the data to constrain the gap properties without being tightly restricted to the specific predictions of evolutionary models.

A schematic representation of our empirical model is provided in Fig.\,\ref{fig:schematic}. The model, $\mathcal{L}$, consists of the background, $\mathcal{L_\mathrm{bg}}$, and gap, $\mathcal{L_\mathrm{gap}}$, components, where the background component represents the smooth colour-magnitude diagram in the absence of the gap. It is given up to the numerically integrated normalisation factor as

\begin{equation}
    \mathcal{L}(\IE, \IE - \HE) \propto \mathcal{L_\mathrm{bg}} \, \mathcal{L_\mathrm{gap}}\,.
    \label{eq:likelihood}
\end{equation}

We modelled the background likelihood, $\mathcal{L_\mathrm{bg}}$, as a Gaussian distribution of colour, $\IE - \HE$, with the central value, $\mu$, width, $\sigma$, and amplitude, $A$, all varying across the CMD. This is expressed as

\begin{equation}
    \mathcal{L_\mathrm{bg}}=\frac{A(\IE, \IE - \HE)}{\sigma(\IE)} \exp\left(-\frac{1}{2}\left[\frac{[\IE - \HE]-\mu(\IE)}{\sigma(\IE)}\right]^2\right)\,.
    \label{eq:likelihood_bg}
\end{equation}

In Eq.\,(\ref{eq:likelihood_bg}), $\mu(\IE)$ and $\sigma(\IE)$ were approximated as second-order polynomials, with coefficients ($\mu_{1-3}$ and $\sigma_{1-3}$) treated as free parameters,

\begin{equation}
    \mu(\IE)=\mu_1 \IE^2 + \mu_2 \IE + \mu_3\,,
    \label{eq:mu}
\end{equation}

\begin{equation}
    \sigma(\IE)=\sigma_1 \IE^2 + \sigma_2 \IE + \sigma_3\,.
    \label{eq:sigma}
\end{equation}

We modelled the gap as a rectangular dip of width $w$ (defined along $\IE$) that follows a straight line in the CMD, parameterized by its central magnitude $g$ and slope $s=\delta \IE/\delta (\IE - \HE)$. We defined the `distance' between a given star and the gap, $\Delta\IE$, according to

\begin{equation}
    \Delta\IE(\IE, \IE - \HE) = \IE - \{g + s [(\IE - \HE) - \mu(g)]\}\,.
    \label{eq:gap_distance}
\end{equation}

The amplitude of the background likelihood, $A$, was modelled as a first-order polynomial function of $\IE$; however, we allowed for a discontinuous change in the polynomial coefficients at the faint edge of the gap (i.e. at $\Delta\IE=w/2$). The functional form of $A$ is given in Eq.\,(\ref{eq:A}), parametrised by $r_\mathrm{below}$, $r_\mathrm{above}$, and $R$,

\begin{equation}
    A(\IE, \IE - \HE)=\begin{cases}
            r_\mathrm{below} (\IE-g-w/2) + R, & \Delta\IE \geq w/2\,, \\
            r_\mathrm{above} (\IE-g-w/2) + 1, & \text{otherwise}\,. \\
        \end{cases}
    \label{eq:A}
\end{equation}
In this parameterization, $R$ is closely related to the size of the `step' feature in the luminosity function discussed earlier. The gap component of the likelihood, $\mathcal{L_\mathrm{gap}}$, is given by

\begin{equation}
    \mathcal{L}_\mathrm{gap}=\begin{cases}
            D, & \text{for } \left|\Delta\IE\right| \leq w/2 \,,\\
            1, & \text{otherwise} \,,\\
        \end{cases}
    \label{eq:likelihood_gap}
\end{equation}
where $D$ is the fractional depth of the gap.

Our likelihood model contains $13$ free parameters: $\mu_1$, $\mu_2$, $\mu_3$, $\sigma_1$, $\sigma_2$, $\sigma_3$, $r_\mathrm{above}$, $r_\mathrm{below}$, $R$, $g$, $w$, $s$, and $D$. We estimated their best-fit values and uncertainties by sampling the parameter space with a Markov chain Monte Carlo (MCMC) approach, using the Goodman--Weare algorithm \citep{MCMC} implemented in the \texttt{emcee} Python package \citep{emcee_2013}. We employed $32$ MCMC walkers and $30\,000$ steps per walker. The best-fit parameters were derived from the walker position with the highest likelihood in the chain, and the errors were taken as standard deviations of the MCMC posteriors.

To evaluate how well our empirical model represents the observed gap, we computed the difference between the logarithmic likelihood of the best-fit model and that of the gap-free ($D=1$), but otherwise identical model. The obtained value, $\Delta\ln\mathcal{L}=33.4$, was compared to its expected distribution, which we estimated by drawing $1000$ synthetic CMDs (with the same sample size as the observed CMD in the range $20<\IE<22$) from the best-fit likelihood model using rejection sampling. To establish the baseline, we also generated $1000$ synthetic datasets from the gap-free likelihood. We inferred the expected distributions of $\Delta\ln\mathcal{L}=48.8\pm 11.0$ and $\Delta\ln\mathcal{L}=-24.4\pm 15.2$ for the best fit and gap-free cases, respectively. The real CMD value of $\Delta\ln\mathcal{L}=33.4$ is well within the distribution predicted by the best-fit model, indicating that our empirical likelihood is consistent with being a good fit to the observed gap within $\approx 1.4\,\sigma$. In contrast, the real CMD is nearly $4$ sigma removed from the $\Delta\ln\mathcal{L}$ range predicted by the gap-free model.

\begin{figure}
    \centering
    \includegraphics[width=\columnwidth]{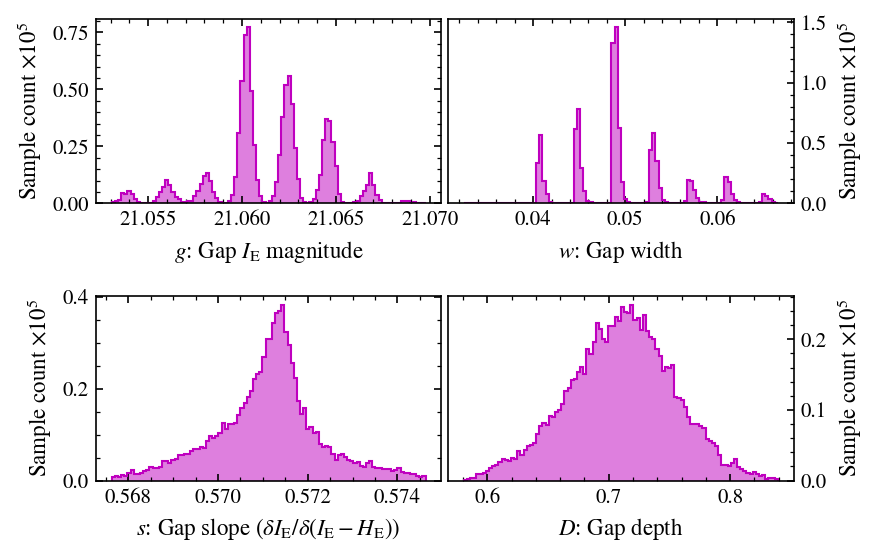}
    \caption{MCMC posterior distributions for the properties of the gap: $\IE$ magnitude of the gap at central colour (\textit{top-left}); gap width in $\IE$ (\textit{top-right}); gap slope in the CMD (\textit{bottom left}); and gap depth, relative to the background distribution of stars in the CMD (\textit{bottom-right}).}
    \label{fig:MCMC_posteriors}
\end{figure}

The best-fit parameters are presented in Table\,\ref{tab:likelihood}. The posterior distributions for the gap parameters (gap magnitude, $g$; gap width, $w$; gap slope, $s$; and gap depth, $D$) are plotted in Fig.\,\ref{fig:MCMC_posteriors}. The posteriors on $g$ and $w$ (upper panels) are multi-modal since the likelihood changes discontinuously with these parameters, as individual stars enter or exit the gap.

\begin{table}[t]
\centering
\caption{Best-fit parameters of our empirical CMD model described in Sect.\,\ref{sec:gap_morphology}, for which the observed CMD of NGC\,6397 near the gap has the highest likelihood.}
\begin{tabularx}{.95\columnwidth}{crll}
\hline
\hline
\noalign{\vskip 2pt}
Parameter & Best fit & Uncertainty & Unit \\ \hline
$\mu_1$ & $-0.0447$ & $\pm0.0012$ & $\mathrm{mag}^{-1}$ \\
$\mu_2$ & $1.993$ & $\pm0.051$ &  \\
$\mu_3$ & $-21.30$ & $\pm0.53$ & $\mathrm{mag}$ \\
$\sigma_1$ & $0.00934$ & $\pm0.00081$ & $\mathrm{mag}^{-1}$  \\
$\sigma_2$ & $-0.386$ & $\pm0.034$ &  \\
$\sigma_3$ & $4.02$ & $\pm0.35$  & $\mathrm{mag}$ \\
$r_\mathrm{above}$ & $0.311$ & $\pm0.022$ & $\mathrm{mag}^{-1}$ \\
$r_\mathrm{below}$ & $-1.22$ & $\pm0.04$ & $\mathrm{mag}^{-1}$ \\
$R$ & $1.327$ & $\pm0.055$ & \\
$g$ & $21.060$ & $\pm0.018$ & $\mathrm{mag}$ \\
$w$ & $0.0486$ & $\pm0.0060$ & $\mathrm{mag}$ \\
$s$ & $0.5713$ & $\pm0.0014$ & \\
$D$ & $0.682$ & $\pm0.044$  \\ \hline
\end{tabularx}
\label{tab:likelihood}
\end{table}

Table\,\ref{tab:likelihood} demonstrates that both $D$ and $R$ differ from unity by between six and seven standard deviations, suggesting that both the gap and the adjacent step in star counts are detected at extremely high statistical confidence. The detection of such fine structures within a GC is uniquely enabled by the \Euclid mission: the combination of high-precision photometry and a wide field of view provides the large statistical sample required to robustly resolve subtle transitions in the CMD and the stellar luminosity function.

\begin{figure}
    \centering
    \includegraphics[width=\columnwidth]{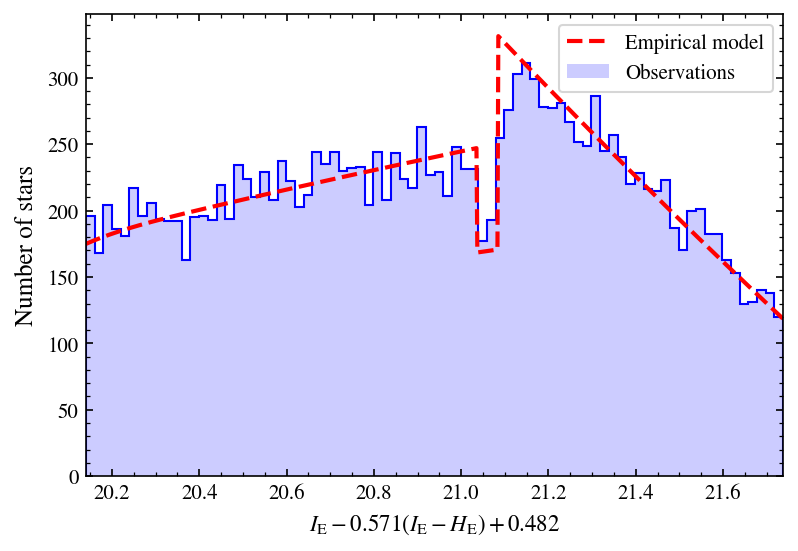}
    \caption{Luminosity function for the transformed $\IE$ magnitude, based on the inferred slope and location of the gap listed in Table\,\ref{tab:likelihood}. The transformation makes the gap horizontal in this space and reveals its true depth and shape in the luminosity space shown in this figure. Our best-fitting model, based on the empirical likelihood in Sect.\,~\ref{sec:gap_morphology}, is also shown.}
    \label{fig:LF_transform}
\end{figure}

The measured CMD slope of the gap, $s$, can be used to derive a linear transformation that makes the gap appear horizontal in the CMD. The luminosity function for $\IE$, transformed in this fashion, is shown in Fig.\,\ref{fig:LF_transform}. Unlike the original luminosity function in the inset of Fig.\,\ref{fig:gap_lf}, the transformed space reveals the true depth and shape of the MS gap.

\subsection{Physical modelling of the gap} \label{sec:gap_physical}

The gap discovered in this study bears a close resemblance to a similar feature in the \textit{Gaia} CMD of Galactic field stars \citep{2018ApJ...861L..11J,Jao_step}, which has been linked to the convective properties of M-dwarfs and, specifically, the production and mixing of $^3\text{He}$ (an intermediate product of the proton-proton chain) as stars develop or dissipate the radiative zone in their interiors \citep{MacDonald_gap_models,Baraffe_gap_models,2021ApJ...907...53F,2021A&A...650A.184M,2023ApJ...944..129B}. While this gap has been well-documented in the solar neighbourhood using \textit{Gaia} and 2MASS data, its detection in a GC has remained elusive until now. Unlike the field population, GC members have much narrower distributions of age, metallicity and distance, which makes the relationship between the gap morphology and the physical properties of the cluster much more direct.

To confirm the physical nature of the gap in NGC\,6397, we produced a new set of stellar evolutionary models using the \texttt{MESA} (\texttt{M}odules for \texttt{E}xperiments in \texttt{S}tellar \texttt{A}strophysics) code, version \texttt{23.05.1} \citep{MESA,MESA_2,MESA_3,MESA_4,MESA_5}. Our modelling setup is described in \cite{roman_6397,roman_47Tuc}. The outer boundary conditions for the stellar structure equations in \texttt{MESA} were set to match the pressures and temperatures derived from a grid of new model atmospheres at the Rosseland optical depth of $100$. Our grid of model atmospheres was calculated using \texttt{BasicATLAS}/\texttt{ATLAS-9}\footnote{The \texttt{ATLAS-9} code offers fast computation of spectra, but we note that it reaches its limit of applicability at the cooler end of the modelled temperature range, where the approximation of plane-parallel atmospheres in local thermodynamic equilibrium becomes less valid, and differences from the pre-computed opacity distribution functions and incompleteness and imprecision of the available molecular line lists become important. While the good agreement between models and observations suggests our \texttt{ATLAS-9} models sufficiently reflect reality, we note the possibility that there may be unaccounted-for systematic errors in the numerical results.} \citep{BasicATLAS,ATLAS5,ATLAS9_1,ATLAS9_2} for effective temperatures, $T_\mathrm{eff}$, between $3500\,\mathrm{K}$ and $5500\,\mathrm{K}$ in steps of $100\,\mathrm{K}$, surface gravities, $\log g$, between $2$ and $5.5$ in steps of $0.25$, metallicities, $[\mathrm{M/H}]$, between $-2.1$ and $-1.7$ in steps of $0.2$, oxygen abundances, $[\mathrm{O/M}]$, between $0$ and $0.6$ in steps of $0.3$, and helium mass fractions, $Y$, between $0.23$ and $0.27$ in steps of $0.02$. A linear interpolation was used to obtain model atmospheres between the grid points.

We first computed the so-called nominal set of evolutionary models for the parameters of NGC\,6397 in the literature. Specifically, we adopted $[\mathrm{M/H}]=-1.9$ from \citet{matteo_NGC6397}, $[\mathrm{O/M}]=0.3$ from \citet{2024A&A...689A..59S}, $Y=0.25$ based on the primordial helium mass fraction from \citet{primordial_Y}, an age of $T=12.6\ \mathrm{Gyr}$ from \citet{matteo_NGC6397}, and the convective mixing length in the interior of $\alpha_\mathrm{MLT}=1.82$ scale heights based on solar calibration \citep{solar_alpha}. In addition to this nominal set of parameters, we computed ten other sets of \texttt{MESA} models, offsetting the parameters listed above one at a time from their nominal values. We considered the offsets in $[\mathrm{M/H}]$ by $\pm 0.2$, offsets in $[\mathrm{O/M}]$ by $\pm 0.3$, offsets in $Y$ by $\pm 0.02$, offsets in $T$ by $\pm1\ \mathrm{Gyr}$ and offsets in $\alpha_\mathrm{MLT}$ by $\pm 0.2$ scale heights.

Near the boundary between fully convective and partly radiative interiors, stars are predicted to oscillate between those two states, driven by the so-called `convective kissing instability' \citep{convective_kissing}. However, in all of our models, the stars in the transition region stop oscillating and settle into stable equilibrium before reaching the target age. We may therefore define a singular stellar mass for each set of \texttt{MESA} models, at which the transition occurs.

For each set of parameters (the nominal set and $10$ perturbations), we computed evolutionary models for stellar masses between $0.3$ and $0.4\,M_\odot$. To accurately reproduce the gap, we first searched for the exact initial mass at which the stellar interior becomes fully convective.\footnote{This search was carried out using Brent's 
method of root finding by requiring the lowest ratio of convective to total flux in the interior of the star to have a very small value (we chose $10^{-50}$).} This transition occurs at $0.354\,M_\odot$ for the nominal parameters of NGC\,6397, and varies by up to $0.01\,M_\odot$ for some of the perturbations. Next, we sampled stellar masses on both sides of the convective transition using adaptive step-size control introduced in \citet{roman_47Tuc}, with the target difference in luminosity between adjacent models of $0.005\ \mathrm{dex}$.\looseness=-1

To convert the physical luminosities predicted by \texttt{MESA} to the observed $\IE$ magnitudes, we utilised the \texttt{synphot} routine of \texttt{BasicATLAS} \citep{BasicATLAS}. In this conversion, we adopted an interstellar reddening of $E(B-V)=0.22$ \citep{matteo_NGC6397} and a distance of $2.458\,\mathrm{kpc}$ \citep{2021MNRAS.505.5957B} as the nominal parameters; however, we also considered offsets by $\pm0.03\ \mathrm{mag}$ in the former, and by $\pm0.12\ \mathrm{kpc}$ in the latter.

We derived a mass-luminosity relationship, $M(\IE)$, from each set of \texttt{MESA} models using first-order spline interpolation. We then combined these relationships with a power-law mass function to derive the model luminosity function of NGC\,6397,

\begin{equation}
        \phi(\IE)\propto \left[M(\IE)\right]^{-\alpha} \left|\frac{\mathrm{d}M(\IE)}{\mathrm{d}\IE}\right|\,.
    \label{eq:LF}
\end{equation}

\begin{figure}
    \centering
    \includegraphics[width=\columnwidth]{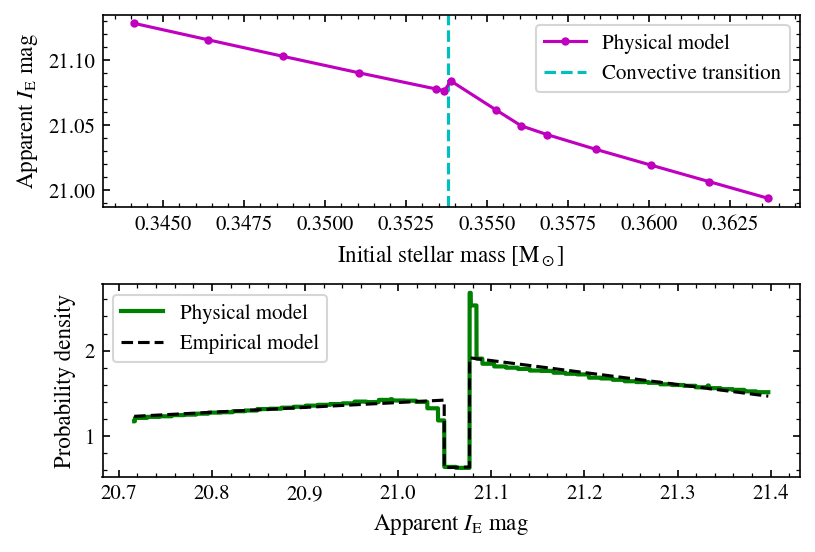}
    \caption{\textit{Top}: Predicted mass-luminosity relationship near the convective transition for the nominal parameters of NGC\,6397 ($[\mathrm{M/H}]=-1.9$, $[\mathrm{O/M}]=0.3$, $Y=0.25$, $T=12.6\ \mathrm{Gyr}$, $\alpha_\mathrm{MLT}=1.82$, $E(B-V)=0.22$, distance of $2.458\,\mathrm{kpc}$, and $\alpha=0$). The points represent individual \texttt{MESA} models. \textit{Bottom}: Corresponding predicted luminosity function. The best-fit empirical model introduced in Sect.\,\ref{sec:gap_morphology} is also shown. The physical model of the luminosity function features a spike in star counts at the faint edge of the gap due to the abrupt discontinuity in the mass-luminosity relationship. It does not matter whether the spike is a genuine feature of the cluster or a modelling artefact, as its range of magnitudes is too narrow to be detected in the observed luminosity function.}
    \label{fig:physical_model}
\end{figure}

Here, $\alpha$ is the power-law index (slope) of the mass function, and $\phi(\IE)$ is the luminosity function (i.e. the probability density of observing a star with the magnitude $\IE$). The choice of $\alpha$ is not important for our purposes, since the gap spans a very narrow range of stellar masses. To demonstrate this, we adopted a flat mass function ($\alpha=0$) as the nominal case, and considered offsets by $\pm 2$, which span the full range of measured values (Fig.\,\ref{fig:mf}). Note that due to the discontinuity of the mass-luminosity relationship near the convective transition, $M(\IE)$ is a multivalued function (i.e. the same $\IE$ magnitude may correspond to two distinct stellar masses). For this reason, we evaluated $\phi(\IE)$ separately on both sides of the discontinuity, and added the results. For the nominal parameters of NGC\,6397, the calculated mass-luminosity relationship and luminosity function are shown in Fig.\,\ref{fig:physical_model}.\looseness=-1

To compare the physical model described in this subsection to the observed properties of the gap, we fitted the empirical likelihood model from Sect.\,\ref{sec:gap_morphology} at the central colour to the nominal physical model of the luminosity function, as well as all perturbations considered in this study (the nominal fit is shown in the bottom panel of Fig.\,\ref{fig:physical_model}). From each fit, we extracted the gap magnitude, gap width, and gap depth ($g$, $w$, and $D$, respectively); as well as the size of the `step' in star counts associated with the gap ($R$). The results of this analysis are shown in Table\,\ref{tab:gap_params}. The best-fit values of these four parameters from Table\,\ref{tab:likelihood} are reported at the bottom of Table\,\ref{tab:gap_params} for convenience.

\begin{table}[t]
\centering
\caption{Predicted and measured gap parameters for NGC\,6397.}
\begin{tabularx}{.95\columnwidth}{lXXXX}
\hline
\hline
\noalign{\vskip 2pt}
Model  & $g$ & $w$ & $D$ & $R$ \\ \hline
Nominal & $21.06$ & $0.03$ & $0.44$ & $1.34$ \\
$[\mathrm{M/H}]=-2.1$ & $20.97$ & $0.03$ & $0.45$ & $1.34$ \\
$[\mathrm{M/H}]=-1.7$ & $21.17$ & $0.02$ & $0.49$ & $1.37$ \\
$[\mathrm{O/M}]=0.0$ & $21.07$ & $0.03$ & $0.44$ & $1.34$ \\
$[\mathrm{O/M}]=0.6$ & $21.07$ & $0.03$ & $0.45$ & $1.34$ \\
$Y=0.23$ & $21.04$ & $0.03$ & $0.49$ & $1.38$ \\
$Y=0.27$ & $21.10$ & $0.03$ & $0.43$ & $1.32$ \\
$T=11.6\ \mathrm{Gyr}$ & $21.07$ & $0.02$ & $0.48$ & $1.36$ \\
$T=13.6\ \mathrm{Gyr}$ & $21.06$ & $0.03$ & $0.43$ & $1.31$ \\
$\alpha_\mathrm{MLT}=1.62$ & $21.06$ & $0.03$ & $0.46$ & $1.34$ \\
$\alpha_\mathrm{MLT}=2.02$ & $21.06$ & $0.03$ & $0.44$ & $1.34$ \\
$\mathrm{dist}=2.338\ \mathrm{kpc}$ & $20.95$ & $0.03$ & $0.44$ & $1.34$ \\
$\mathrm{dist}=2.578\ \mathrm{kpc}$ & $21.17$ & $0.03$ & $0.44$ & $1.34$ \\
$E(B-V)=0.19$ & $21.00$ & $0.03$ & $0.45$ & $1.33$ \\
$E(B-V)=0.25$ & $21.13$ & $0.03$ & $0.44$ & $1.33$ \\
$\alpha=-2$ & $21.06$ & $0.03$ & $0.44$ & $1.31$ \\
$\alpha=+2$ & $21.06$ & $0.03$ & $0.44$ & $1.37$ \\ \hline
Measured & $21.060$ & $0.0486$ & $0.682$ & $1.327$ \\
 & $\pm 0.018$ & $\pm 0.0060$ & $\pm 0.044$ & $\pm 0.055$ \\ \hline
\end{tabularx}
\tablefoot{
$g$ is the predicted magnitude, $w$ is the width, $D$ is the depth of the gap, and $R$ is the size of the associated step in star counts. 
The nominal model assumes $[\mathrm{M/H}]=-1.9$, $[\mathrm{O/M}]=0.3$, $Y=0.25$, $T=12.6\ \mathrm{Gyr}$, $\alpha_\mathrm{MLT}=1.82$, $E(B-V)=0.22$, $d=2.458\ \mathrm{kpc}$, and $\alpha=0$. 
The remaining rows show the effect of varying one parameter at a time. 
Measured values at the bottom are taken from Table\,\ref{tab:likelihood}.
}
\label{tab:gap_params}
\end{table}

Comparing the measured and predicted parameters of the gap, we note the remarkable agreement in $g$ and $R$ for the nominal parameters of NGC\,6397. Interestingly, the physical model under-predicts the width $w$ and depth $D$ of the gap, while maintaining excellent agreement with the integrated area of the gap, $w\,(1-D)$. While this discrepancy may be indicative of model deficiencies (e.g. a failure to fully resolve the convective kissing instability due to finite time stepping), the agreement in the integrated gap area hints at an alternative interpretation: the observed gap may correspond to a broadened version of the theoretical prediction, effectively arising from a convolution with some probability density distribution. Such a scenario would occur if the observed CMD represents a superposition of gaps associated with a spread in a parameter that controls the gap magnitude ($g$), most plausibly $[\mathrm{M/H}]$ (see Table\,\ref{tab:gap_params}). In this case, the width of the gap ($w$) may encode the intrinsic dispersion of this parameter within NGC\,6397. We defer a more detailed exploration of this interpretation to future work.

The variations in $w$, $D$, and $R$ across all sets of parameters considered in this study are smaller or comparable with the measurement errors in those parameters for our sample size. The magnitude of the gap ($g$) depends strongly on metallicity ($g$ gets fainter by about $0.05\,\mathrm{mag}$ per $0.1\,\mathrm{dex}$ increase in $[\mathrm{M/H}]$), in addition to reddening and distance to the cluster. In contrast, the effect of the helium mass fraction on the gap magnitude is much weaker ($g$ gets fainter by about $0.015\,\mathrm{mag}$ per $0.01$ increase in $Y$), and the effect of other parameters is undetectably small.

If the reddening and average metallicity of NGC\,6397 can be accurately inferred from independent observables, then the gap magnitude can be used as a `standard candle' to estimate the distance to the cluster. For example, if we adopt the average metallicity and reddening of NGC\,6397 from \citet{matteo_NGC6397} ($[\mathrm{M/H}]=-1.88\pm0.04$, $E(B-V)=0.220\pm0.015$), then we can use Table\,\ref{tab:gap_params} to estimate the absolute $\IE$ magnitude of the gap as $9.120\pm 0.037$. Comparing it to the observed apparent magnitude of the gap in Table\,\ref{tab:likelihood}, we may infer the distance to NGC\,6397 as $2443\pm46\ \mathrm{pc}$. This estimate has a better formal precision than the equivalent measurement from \textit{Gaia} parallaxes ($2458\pm60\ \mathrm{pc}$, \citealt{2021MNRAS.505.5957B}). In practice, the distance uncertainty is likely underestimated due to zero-point offsets in the photometric calibration, errors in our physical model, systematic errors in $[\mathrm{M/H}]$ and $E(B-V)$, as well as biases due to intrinsic metallicity dispersion of the cluster and differential reddening.\looseness=-1

Since NGC\,6397 has a very small dispersion in $Y$ ($\sim 0.01$, \citealt{6397_small_Y_spread}), the $[\mathrm{M/H}]$ dispersion of the cluster must be the dominant contributor to the observed slope of the gap $s$. Here, $[\mathrm{M/H}]$ encompasses all metals, including light elements whose large star-to-star variations are the defining signature of multiple stellar populations in GCs \citep[e.g.][]{2018ARA&A..56...83B,2020CassimPOPsRv}. However, at the extremely low metallicity of NGC\,6397, the influence of oxygen -- the dominant metal and typically a major driver of atmospheric opacity -- is strongly suppressed. Our models indicate that even large variations of an order of $\sim0.6$ dex in $[\mathrm{O/M}]$ produce negligible changes in the gap magnitude (see Table\,\ref{tab:gap_params}). 
While we have not explicitly modelled variations in other light elements, NGC\,6397 is known to host significant spreads in nitrogen \citep{Carretta2005, Lind2011}, which might also influence atmospheric opacities, albeit to a lower extent than oxygen \citep{VandenBerg2012}. Furthermore, the spread in helium could be larger than the average $\Delta Y \approx 0.01$ reported by \cite{6397_small_Y_spread}. However, our tests (Table\,\ref{tab:gap_params}) demonstrate that even a substantial 0.6 dex spread in $[\mathrm{O/M}]$ -- a more significant opacity source -- fails to shift the gap parameters. This hints at a more complex combination of factors as a perhaps more likely cause for changes in the morphology of the gap, requiring more detailed modelling that is beyond the scope of this work.

For the best-fit parameters in Table\,\ref{tab:likelihood}, it can be calculated that the centre of the gap spans $\IE$ magnitudes between $21.04$ and $21.08$ within one standard deviation of the colour spread. This corresponds to $0.045\,\mathrm{dex}$ of 1\,$\sigma$ spread in $[\mathrm{M/H}]$. This estimate has a negligible random error ($\approx 0.0003\ \mathrm{dex}$) and its systematic error budget is almost entirely dominated by the accuracy of the $[\mathrm{M/H}]$--$g$ relationship predicted by the physical model. We also note that a slope in the gap would emerge only if $[\mathrm{M/H}]$ is \mbox{(anti-) correlated} with the photometric colour of the star (otherwise, $[\mathrm{M/H}]$ dispersion would broaden the gap without tilting it). The colour spread is often associated with the scatter in light-element abundances (i.e. multiple populations). Our result therefore implies a degree of correlation between $[\mathrm{M/H}]$ and light elements.

\section{Conclusions}

In this work, we have introduced and validated a multiple-pass photometric and astrometric data reduction pipeline tailored for the \Euclid space telescope. By performing iterative PSF fitting directly on individual exposures and leveraging effective PSF models, our method bypasses the precision loss inherent in traditional image resampling and `single-pass’ catalogues. Through extensive artificial-star tests, we quantified the completeness levels and systematic errors of our data set as a function of stellar density and magnitude. These tests confirm that our multiple-pass routine maintains high recovery rates across the cluster; specifically, while the $50\%$ completeness level reaches $\IE = 26.2$ and $m_{\text{NISP}} = 24.4$ in the lower density outskirts, this limit drops to $\IE = 24.6$ and $m_{\text{NISP}} = 23.2$ in the innermost core regions due to extreme stellar crowding.

Our analysis of the globular cluster NGC~6397 serves as a proof-of-concept for the potential of \Euclid in the field of resolved stellar populations. Our multiple-pass approach significantly improves source deblending and photometric precision in the crowded regions surrounding NGC~6397. We coupled \Euclid data with multi-epoch HST observations to derive high-precision proper motions, enabling a detailed characterisation of the internal kinematics of the cluster and the degree of energy equipartition as a function of distance from the centre. Specifically, we found that the degree of energy equipartition does not vary significantly with distance from the centre, nor is there a significant difference between the radial and tangential components of the velocity dispersion. This uniformity indicates that NGC\,6397 is a dynamically old system that has reached an advanced evolutionary state where such variations have been smoothed out, in general agreement with theoretical expectations for clusters in advanced stages of evolution \citep[e.g.][]{2013MNRAS.435.3272T}.

The wide field of view provided by \Euclid further allowed us to study the present-day local mass function (MF) and binary fraction from the inner regions to the outskirts, revealing a clear signature of mass segregation. This is characterised by a significant flattening of the MF in the core and a progressive steepening in the external regions, where the global MF slope suggests the cluster MF has been significantly flattened by the preferential loss of low-mass stars due to two-body relaxation and the Galactic tidal field \citep{1997MNRAS.289..898V,2003MNRAS.340..227B}.
The analysis of the radial distribution of the binary fraction is clearly unimodal, peaking in the core and decreasing towards the outskirts. Consistently with recent findings (\citealt{2026arXiv260207119B,Cadelano2026}), this trend reflects the combined influence of dynamical evolution, binary disruption, and the distinct spatial distributions of multiple populations. Such a monotonic decrease is the expected signature for a dynamically evolved, post-core-collapse cluster such as NGC\,6397.\looseness=-1

Finally, our multiple-pass, high-precision CMD revealed a subtle under-density that is compatible with the transition of stellar interiors from having a radiative zone to full convection. This feature has been previously observed in the CMD of field stars; however, we report the first detection of the gap in a GC. We constructed a fully self-consistent physical model of the luminosity function that exhibits remarkable agreement with the observed properties of the gap. We also determined that the magnitude of the gap can be used as a `standard candle' to refine distance estimates to GCs, while the slope of the gap in the CMD might offer the most precise estimate of the intrinsic member-to-member metallicity dispersion in the cluster. We estimated the 1\,$\sigma$ metallicity dispersion of NGC\,6397 as $0.045\,\mathrm{dex}$ with a negligible random error. We defer the study of the systematic errors associated with those measurements to a future study aimed at conducting a critical investigation of the accuracy of our stellar models.

Ultimately, the detection of such subtle features underscores that the wide field of view of \Euclid, combined with the precision enabled by our multiple-pass routine, offers an unprecedented opportunity to study Galactic globular clusters in their entirety. While HST provides unparalleled depth in small apertures, \Euclid provides the robust statistics necessary to identify subtle atmospheric transitions that are often lost in the more localised footprints of other instruments. Moreover, with the acquisition of multiple epochs, \Euclid could enable high-precision proper motions and kinematic analyses across the entire field of view \citep{2025A&A...704A.193B}. 
This would significantly improve cluster-field decontamination in the outskirts, where the density of cluster members is comparable to the foreground and background populations. Such a comprehensive data set would allow for a more rigorous investigation of the cluster's tidal tails and the overall impact of the Galactic gravitational field on its long-term dynamical evolution.

\begin{acknowledgements}
The authors wish to warmly thank the referee, Nate Bastian, and the editor, Thierry Forveille, for their constructive comments and suggestions that improved the quality of this work.
\AckEC
\AckERO
Based on observations with the NASA/ESA \textit{Hubble} Space Telescope obtained from the Data Archive at the Space Telescope Science Institute (STScI), which is operated by the Association of Universities for Research in Astronomy, Incorporated, under NASA contract NAS5-26555.
This work has made use of data from the European Space Agency (ESA) mission Gaia (\url{https://www.cosmos.esa.int/gaia}), processed by the Gaia Data Processing and Analysis Consortium (DPAC, \url{https://www.cosmos.esa.int/web/gaia/dpac/consortium}). Funding for the DPAC has been provided by national institutions, in particular the institutions participating in the Gaia Multilateral Agreement.
MG acknowledges support from STScI DRF grant D0101.90373. ED acknowledges financial support from the INAF Data analysis Research Grant (PI E. Dalessandro) of the ``Bando Astrofisica Fondamentale 2024''.

\end{acknowledgements}

\bibliography{bibliography} 

\appendix

\section{Artificial stars}
\label{sec:artificial}

Artificial stars (ASs) are essential to quantify the completeness and photometric accuracy of our measurements, particularly in crowded fields \citep[see e.g.][]{2009ApJ...697..965B}. The \texttt{KS2} code allows us to perform AS tests by injecting a sample of synthetic stars of known input magnitude and position into the images one at a time, which are then reprocessed blindly with the same input parameters used for the real measurements. ASs are added, measured, and removed, so that each AS never interferes with the others. 

We generated one million ASs on a fiducial line that traces the main sequence of NGC\,6397. By comparing the recovered magnitudes ($m_{\rm out}$) of the ASs to their known input magnitudes ($m_{\rm in}$) and positions, the recovery fraction (completeness) and the magnitude bias ($\Delta_m=m_{\rm in}-m_{\rm out}$) can be systematically mapped as a function of magnitude and position on the detector. We consider an AS to be successfully recovered in a given filter if it is found within 0.5\,pixels in both coordinates and within 0.75\,mag of its input value, as in \cite{2017ApJ...842....6B}. These results are crucial for establishing the observational limits of the data and for correcting observed luminosity functions for incompleteness effects. At this stage, we did not apply any quality cut to the ASs.

\subsection{Input-output systematic errors}\label{sec:io}

\begin{figure}
    \centering
    \includegraphics[width=\columnwidth]{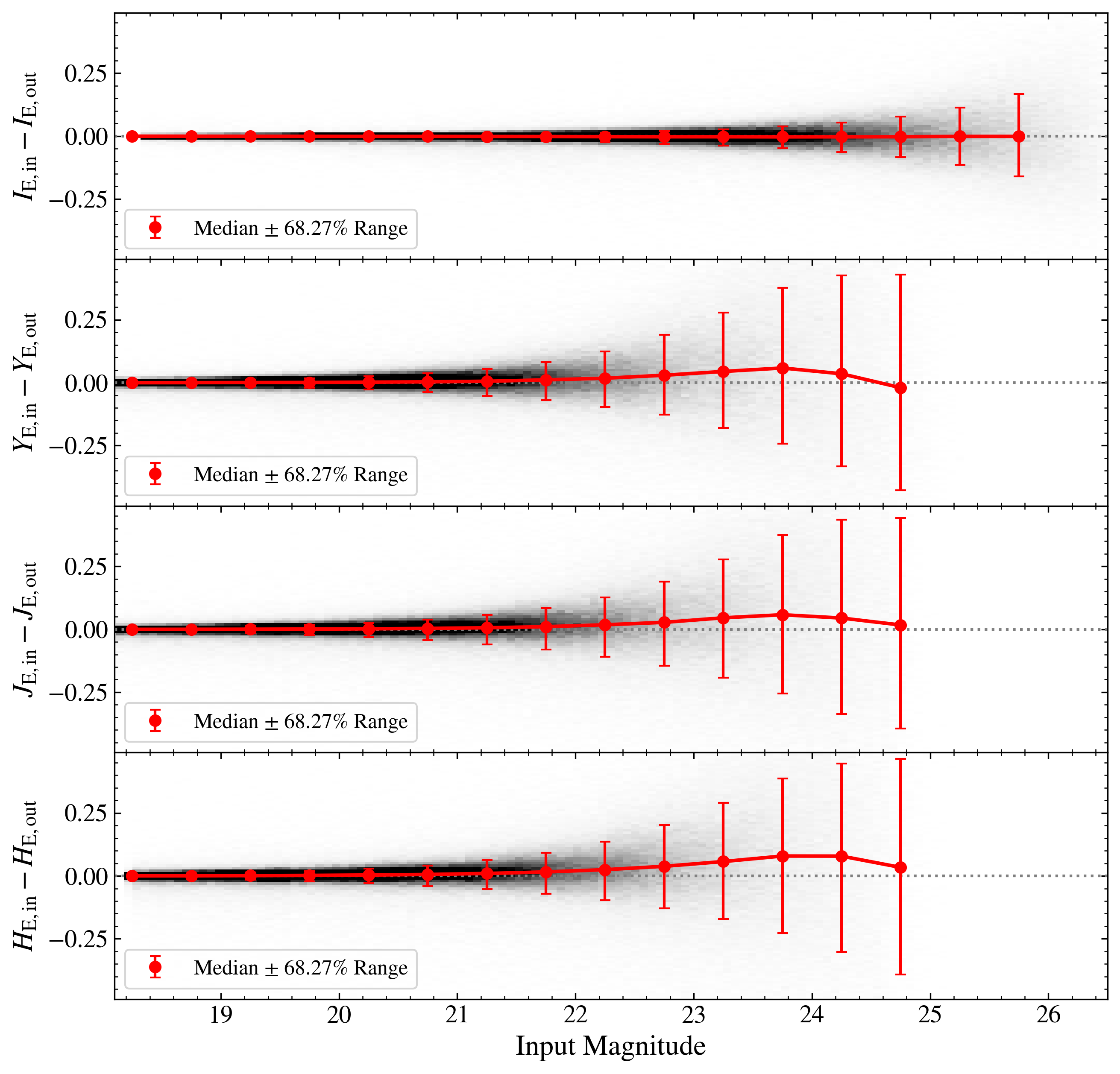}
    \caption{Distribution of photometric residuals ($\Delta m=m_{\rm in}-m_{\rm out}$) as a function of input magnitude for the VIS and NISP filters, measured with \texttt{KS2} method 1. The red dots indicate the median bias calculated in 0.5 mag bins. Error bars represent the 68.27th percentile in each magnitude bin. No quality cuts have been applied.
    }
    \label{fig:as_res_m1}
\end{figure}

\begin{figure}
    \centering
    \includegraphics[width=\columnwidth]{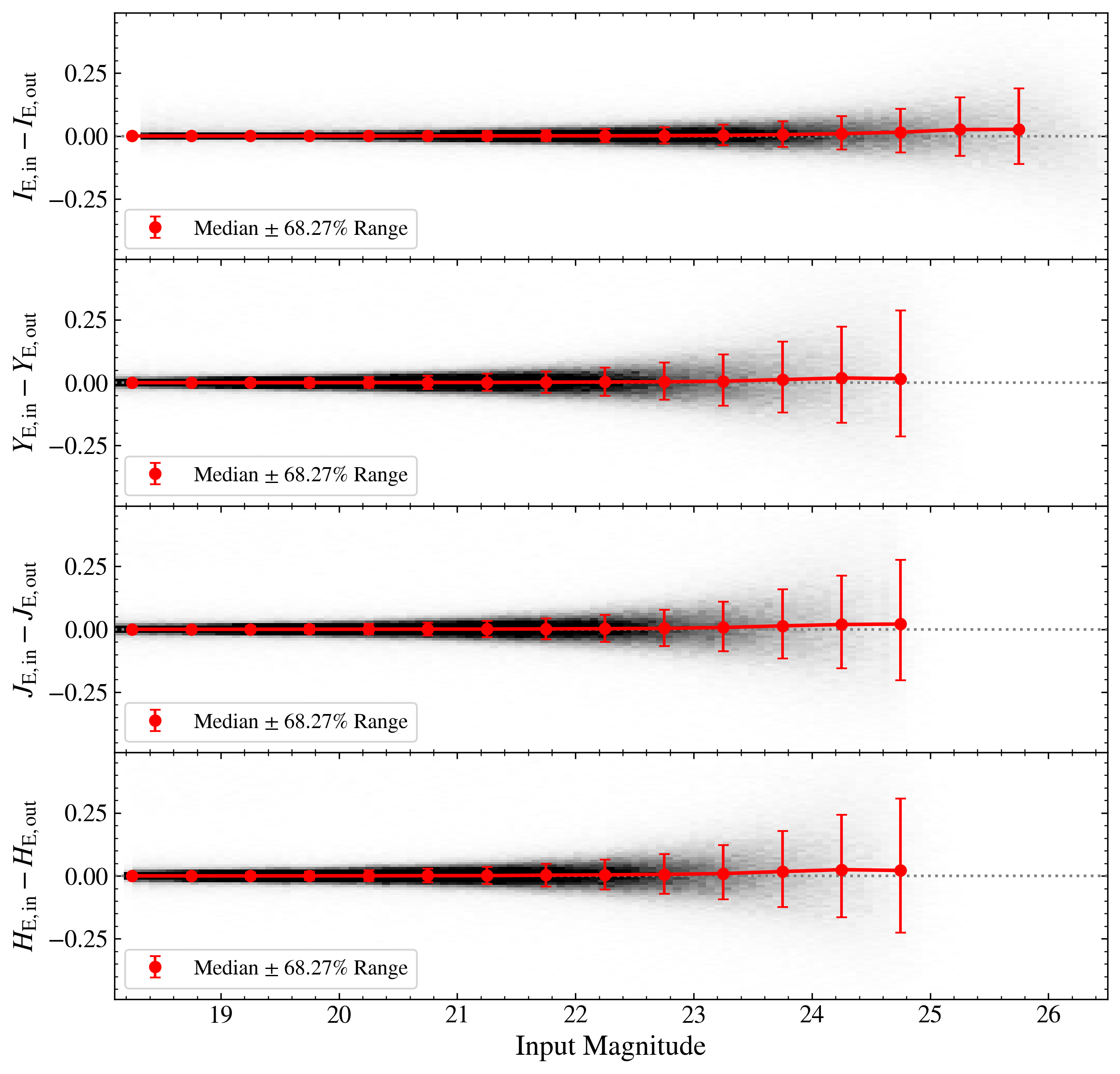}
    \caption{Similar to Fig.\,\ref{fig:as_res_m1}, but for \texttt{KS2} method 2 photometry.}
    \label{fig:as_res_m2}
\end{figure}

To characterize the photometric performance and quantify systematic uncertainties across the \Euclid passbands, we analyse the residuals between the input and recovered magnitudes of artificial stars. Figure\,\ref{fig:as_res_m1} displays the magnitude bias as a function of the input magnitude for the VIS $\IE$ and NISP $\YE$, $\JE$, and $\HE$ filters. In each panel, the red dots indicate the median residual calculated within each 0.5\,mag bin, while the associated error bars represent the 68.27th percentile of the residual distribution. The AS tests allow us to estimate the photometric errors and potential systematic offsets at each magnitude level. We performed this analysis for both \texttt{KS2} methods (Figs.\,\ref{fig:as_res_m1} and \ref{fig:as_res_m2}). A comparison of the two figures shows that method 2 provides a more accurate recovery with reduced scatter, particularly for faint stars. This improvement is especially pronounced in the NISP filters; given their shorter exposure times and lower angular resolution compared to VIS, the NISP photometry benefits significantly from using fixed positions derived from VIS.

\subsection{Completeness}

\begin{figure}
    \centering
    \includegraphics[width=\columnwidth]{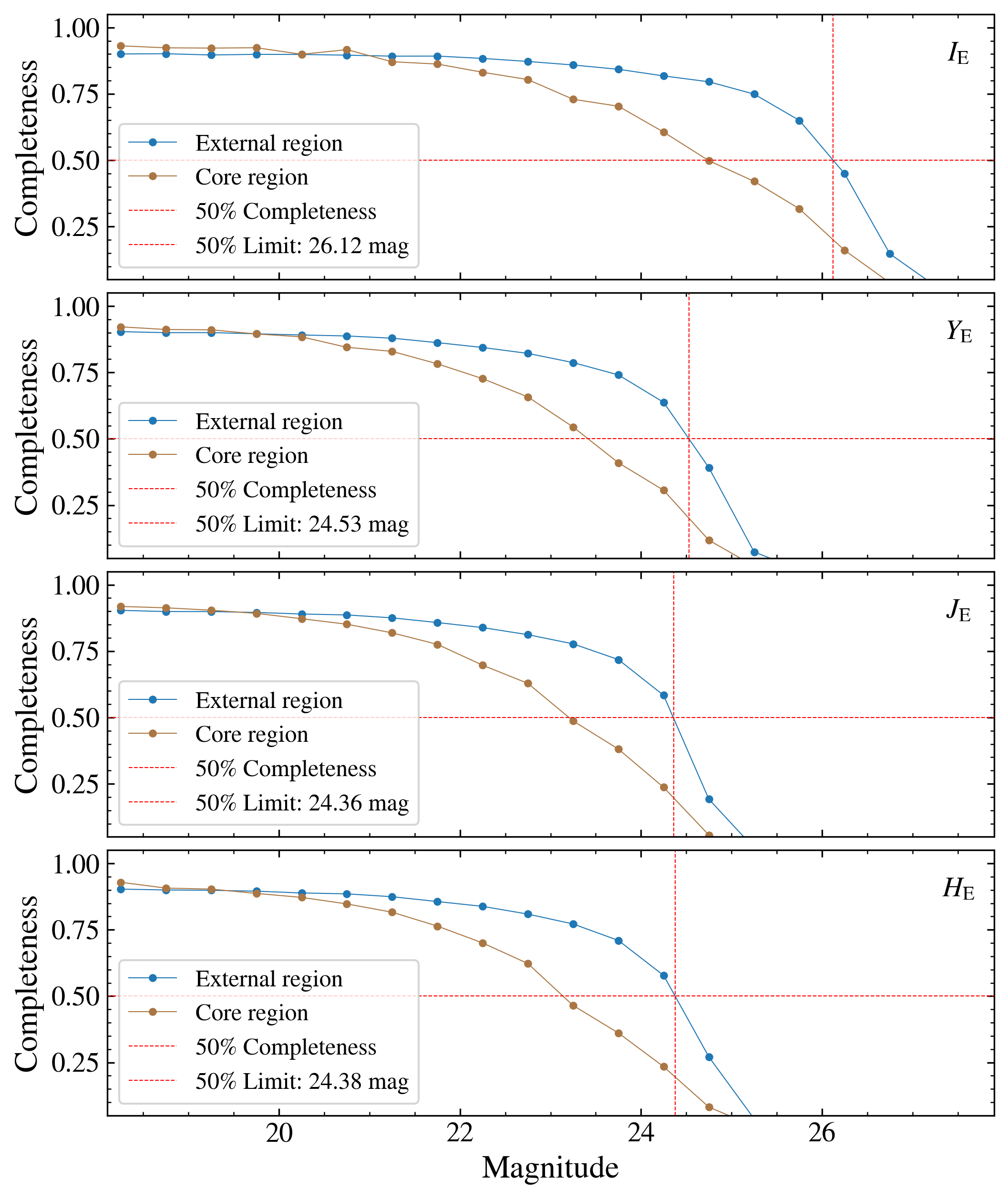}
    \caption{Completeness as a function of input magnitude for the VIS (\textit{top}) and NISP (\textit{bottom}) filters. Blue and red curves represent the recovery fraction in the external regions and the core region of the cluster, respectively. The horizontal dashed line indicates the 50\% completeness level. In the outskirts of the cluster, the 50\% threshold is reached at $m \approx 26$ for VIS and $m \approx 24$ for NISP, while crowding in the core shifts these limits to about 1\,mag brighter. No quality cuts have been applied.
    }
    \label{fig:compl_1d}
\end{figure}

\begin{figure}
    \centering
    \includegraphics[width=\columnwidth]{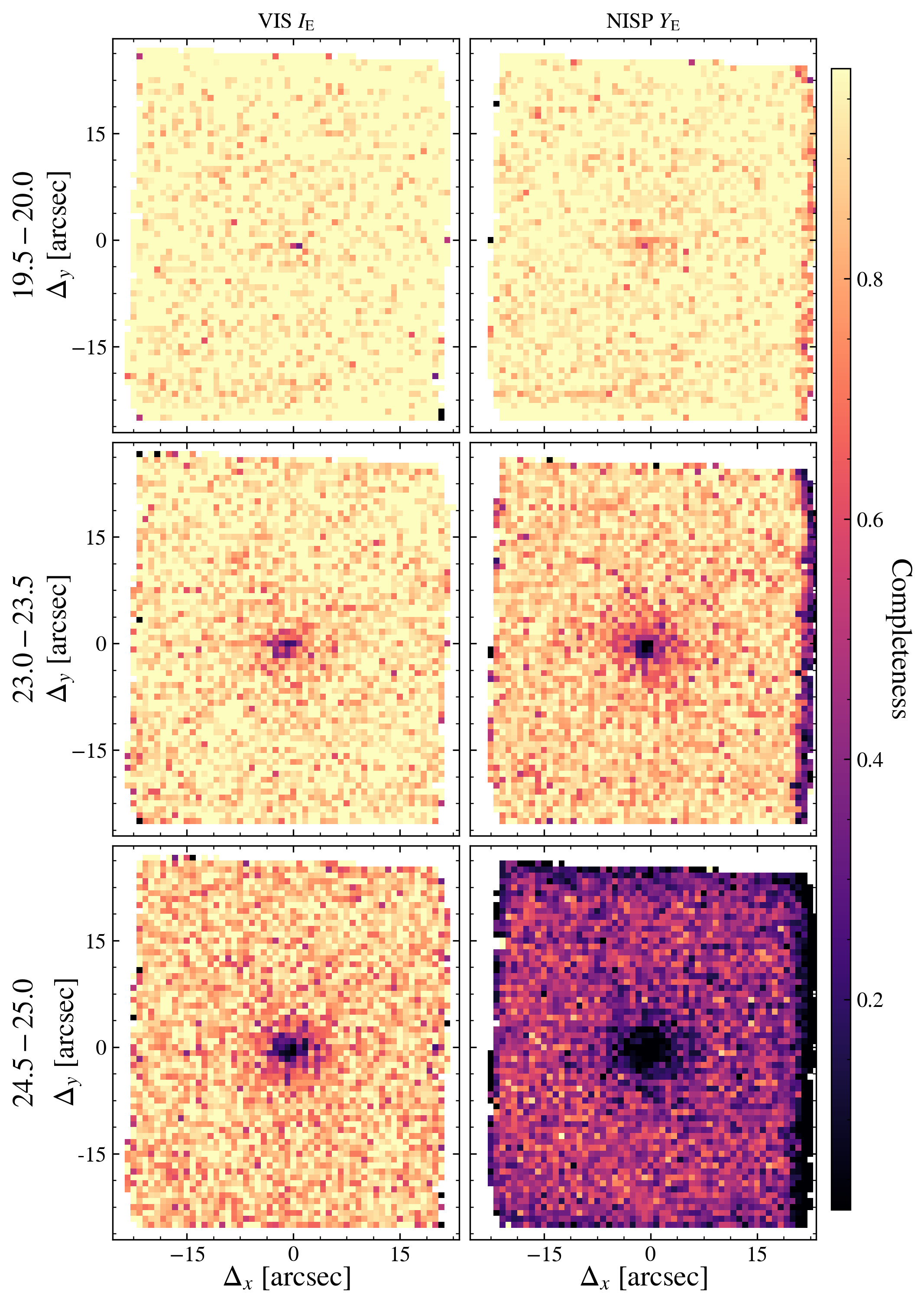}
    \caption{2D completeness maps of the \Euclid NGC\,6397 field of view. The maps are derived by dividing the field into a $500\times 500$ pixel grid and calculating the recovery fraction of artificial stars within each cell. Columns display results for the VIS $\IE$ (\textit{left}) and NISP $\YE$ (\textit{right}) filters. From top to bottom, the rows correspond to increasingly fainter magnitude intervals: [19.5, 20.0]; [23.0, 23.5]; and [24.5, 25.0]. The colour scale indicates the recovery fraction. The clear radial gradient, particularly evident in the fainter magnitude bins, illustrates the severe impact of stellar crowding and elevated background levels on detection efficiency within the cluster's inner regions. No quality cuts have been applied.}
    \label{fig:compl_2d}
\end{figure}

The recovery fraction of ASs enables us to empirically determine the completeness of our data set as a function of both stellar magnitude and local density. Figure\,\ref{fig:compl_1d} presents the one-dimensional completeness curves for each \Euclid filter. To highlight the impact of crowding on our detection efficiency, we compare the recovery rates in two distinct environments: the dense inner region of the cluster (red curves); and the sparser external region (blue curves). Given the core-collapsed nature of NGC\,6397, we define the inner region as the area within $1.5\,r_{\rm h}$, where $r_{\rm h}=2^\prime\!.9$ is the half-light radius \citep[][2010 edition]{1996AJ....112.1487H}, a threshold chosen to capture the high-density environment where crowding and blending are most pronounced, while maintaining sufficient statistics for the AS analysis.

In the external region, the 50\% completeness level is reached at $\IE\approx$\,26\,mag and at $m\approx$\,24\,mag in NISP filters. However, in the central region, the crowding and elevated background levels significantly hamper source detection, resulting in a 50\% recovery threshold that is approximately 1\,mag brighter compared to the outskirts, across all passbands. 

We derived 2D completeness maps across the entire field of view to capture spatial variations in detection efficiency. We divided the field into a grid of $500\times 500$ pixels and computed the local recovery fraction within each cell. Figure\,\ref{fig:compl_2d} displays three representative slices of these 2D maps, with the first column representing the $\IE$ passband and the second column showing $\YE$. Each row corresponds to a distinct magnitude interval. These maps clearly demonstrate the impact of crowding, where the completeness drops abruptly in the high-density inner regions.

\end{document}